\documentclass[twocolumn]{aastex62}
\pdfoutput=1
\usepackage{amsmath,amstext,multirow}
\usepackage[T1]{fontenc}
\usepackage{apjfonts,xcolor}
\usepackage[figure,figure*]{hypcap}
\usepackage{enumitem}

\def\be{\begin{eqnarray}}   \def\ee{\end{eqnarray}}
\def\ben{\begin{equation}\begin{aligned}} \def\een{\end{aligned}\end{equation}}
\shorttitle{Abundance relations}
\shortauthors{Sharma et al.}
\newcommand{\teff}{$T_{\rm{eff}}$}

\newcommand{\numax}{$\nu_{\rm{max}}$}
\newcommand{\dnu}{$\Delta\nu$}

\begin{document}

\title{The GALAH Survey: Dependence of elemental abundances on age and metallicity for stars in the Galactic disc}

\author[0000-0002-0920-809X]{Sanjib~Sharma}
\affiliation{Sydney Institute for Astronomy, School of Physics, A28, The University of Sydney, NSW 2006, Australia}
\affiliation{Centre of Excellence for Astrophysics in Three Dimensions (ASTRO-3D), Australia}
\author[0000-0001-7294-9766]{Michael~R.~Hayden}
\affiliation{Sydney Institute for Astronomy, School of Physics, A28, The University of Sydney, NSW 2006, Australia}
\affiliation{Centre of Excellence for Astrophysics in Three Dimensions (ASTRO-3D), Australia}
\author[0000-0001-7516-4016]{Joss~Bland-Hawthorn}
\affiliation{Sydney Institute for Astronomy, School of Physics, A28, The University of Sydney, NSW 2006, Australia}
\affiliation{Centre of Excellence for Astrophysics in Three Dimensions (ASTRO-3D), Australia}
\author{Dennis~Stello}
\affiliation{School of Physics, University of New South Wales, Sydney, NSW 2052, Australia}
\affiliation{Stellar Astrophysics Centre, Department of Physics and Astronomy, Aarhus University, DK-8000 Aarhus C, Denmark}
\affiliation{ARC Centre of Excellence for All Sky Astrophysics in Three Dimensions (ASTRO-3D)}
\author[0000-0002-4031-8553]{Sven~Buder}
\affiliation{Research School of Astronomy \& Astrophysics, Australian National University, ACT 2611, Australia}
\affiliation{Centre of Excellence for Astrophysics in Three Dimensions (ASTRO-3D), Australia}
\author[0000-0002-7550-7151]{Joel~C.~Zinn}
\affiliation{School of Physics, University of New South Wales, Sydney, NSW 2052, Australia}
\altaffiliation{NSF Astronomy and Astrophysics Postdoctoral Fellow.}
\affiliation{Department of Astrophysics, American Museum of Natural History, Central Park West at 79th Street, NY 10024, USA}
\affiliation{School of Physics, University of New South Wales, Barker Street, Sydney, NSW 2052, Australia}
\author{Lorenzo~Spina}
\affiliation{School of Physics and Astronomy, Monash University, VIC 3800, Australia}
\author{Thomas~Kallinger}
\affiliation{Institute of Astrophysics, University of Vienna, Türkenschanzstrasse 17, Vienna 1180, Austria}
\author{Martin~Asplund}
\affiliation{Max Planck Institute for Astrophysics, Karl-Schwarzschild-Str. 1, D-85741 Garching, Germany}
\author[0000-0001-7362-1682]{Gayandhi~M.~De~Silva}
\affiliation{Australian Astronomical Optics, Faculty of Science and Engineering, Macquarie University, Macquarie Park, NSW 2113, Australia}
\affiliation{Macquarie University Research Centre for Astronomy, Astrophysics \& Astrophotonics, Sydney, NSW 2109, Australia}
\author[0000-0002-2662-3762]{Valentina~{D'Orazi}}
\affiliation{Istituto Nazionale di Astrofisica, Osservatorio Astronomico di Padova, vicolo dell'Osservatorio 5, 35122, Padova, Italy}
\author[0000-0001-6280-1207]{Ken~C.~Freeman}
\affiliation{Research School of Astronomy \& Astrophysics, Australian National University, ACT 2611, Australia}
\author{Janez~Kos}
\affiliation{Faculty of Mathematics and Physics, University of Ljubljana, Jadranska 19, 1000 Ljubljana, Slovenia}
\author[0000-0003-3081-9319]{Geraint~F.~Lewis}
\affiliation{Sydney Institute for Astronomy, School of Physics, A28, The University of Sydney, NSW 2006, Australia}
\author{Jane~Lin}
\affiliation{Research School of Astronomy \& Astrophysics, Australian National University, ACT 2611, Australia}
\affiliation{Centre of Excellence for Astrophysics in Three Dimensions (ASTRO-3D), Australia}
\author{Karin~Lind}
\affiliation{Department of Astronomy, Stockholm University, AlbaNova University Centre, SE-106 91 Stockholm, Sweden}
\author[0000-0002-3430-4163]{Sarah~L.~Martell}
\affiliation{School of Physics, UNSW, Sydney, NSW 2052, Australia}
\affiliation{Centre of Excellence for Astrophysics in Three Dimensions (ASTRO-3D), Australia}
\author[0000-0003-0110-0540]{Katharine J. Schlesinger}
\affiliation{Research School of Astronomy \& Astrophysics, Australian National University, ACT 2611, Australia}
\author[0000-0002-8165-2507]{Jeffrey~D.~Simpson}
\affiliation{School of Physics, UNSW, Sydney, NSW 2052, Australia}
\affiliation{Centre of Excellence for Astrophysics in Three Dimensions (ASTRO-3D), Australia}
\author[0000-0003-1124-8477]{Daniel~B.~Zucker}
\affiliation{Department of Physics and Astronomy, Macquarie University, Sydney, NSW 2109, Australia}
\affiliation{Macquarie University Research Centre for Astronomy, Astrophysics \& Astrophotonics, Sydney, NSW 2109, Australia}
\author[0000-0002-2325-8763]{Toma\v{z}~Zwitter}
\affiliation{Faculty of Mathematics and Physics, University of Ljubljana, Jadranska 19, 1000 Ljubljana, Slovenia}
\author{Boquan~Chen}
\affiliation{Sydney Institute for Astronomy, School of Physics, The University of Sydney, NSW 2006, Australia}
\author{Klemen~Cotar}
\affiliation{Faculty of Mathematics and Physics, University of Ljubljana, Jadranska 19, 1000 Ljubljana, Slovenia}
\author{Prajwal~R.~Kafle}
\affiliation{International Centre for Radio Astronomy Research (ICRAR), The University of Western Australia, 35 Stirling Highway, \\Crawley, WA 6009, Australia}
\author{Shourya~Khanna}
\affiliation{University of Groningen, 9712 CP, Groningen, Netherlands}
\author{Purmortal~Wang}
\affiliation{Sydney Institute for Astronomy, School of Physics, The University of Sydney, NSW 2006, Australia}
\author{Rob~A.~Wittenmyer}
\affiliation{Centre for Astrophysics, University of Southern Queensland, Toowoomba, Queensland 4350, Australia}

\begin{abstract}
Using data from the GALAH survey, we explore the dependence of elemental abundances on stellar age and metallicity among Galactic disc stars. We find that the abundance of most elements can be predicted from age and [Fe/H] with an intrinsic scatter of about 0.03 dex. We discuss the possible causes for the existence of the abundance-age-metallicity relations.
Using a stochastic chemical enrichment scheme based on the size of Supernovae remnants, we show the intrinsic scatter is expected to be small, about 0.05 dex or even smaller if there is additional mixing in the ISM. Elemental abundances show trends with both age and metallicity and the relationship is well described by a simple model in which the dependence of abundance ([X/Fe]) on age and [Fe/H] are additively separable. Elements can be grouped based on the  direction of their abundance gradient in the (age,[Fe/H]) plane and different groups can be roughly associated with three distinct nucleosynthetic production sites, the exploding massive stars, the exploding white dwarfs and the AGB stars.  However, the abundances of some elements, like Co, La, and Li, show large scatter for a given age and metallicity, suggesting processes other than simple Galactic chemical evolution are at play.  We also compare the abundance trends of main-sequence turn-off stars against that of giants, whose ages were estimated using asteroseismic information from the K2 mission. For most elements, the trends of main-sequence turn-off stars are similar to that of giants.  The existence of abundance relations implies that we can estimate the age and birth radius of disc stars, which is important for studying the dynamic and chemical evolution of the Galaxy.
\end{abstract}

\section{Introduction}

Understanding the formation and evolution of the Milky Way is a major challenge for modern astronomy.  Importantly, stellar age and composition can reveal the key aspects of the evolution in this regard.  The usefulness of the chemical composition stems from three key facts \citep{2009nceg.book.....P}.
Firstly, except for a few elements like H, He, Li and B which are made by the Big Bang nucelosynthesis, all elements are synthesized in the interiors of stars and are released into the surrounding gas thereby enriching it when stars die. Successive generation of stars are born out of gas enriched by previous generations, which means chemical information is passed down from one generation of stars to the next.  Secondly, although stars synthesize new elements in their interior, their surface composition remains, with some exceptions, largely unchanged and this is what we observe and measure.
Hence, the surface chemical composition of stars encode the information about the environment from which they were born \citep{2002ARA&A..40..487F}.  Finally, there are various production sites for chemical elements (such as core collapse supernovae, explosion of white dwarfs in binary systems, and asymptotic giant branch stars) and the nucleosynthetic yields of elements varies from one site to another. The occurrence rate of the various sites change with time and location. This means that the information carried by the different elements is not the same but is rich and useful when considered as an ensemble.

There is a rich history of chemical evolution models, trying to predict the distribution of age and elemental abundances for stars in the Galaxy, including analytical models, semi-analytical models, and cosmological simulations \citep{1997ApJ...477..765C,2009MNRAS.396..203S,2015A&A...580A.126K,2013A&A...558A...9M, 2017ApJ...835..224A, 2018MNRAS.477.5072M, 2020MNRAS.491.5435B, 2020ApJ...900..179K}.
However, understanding chemical enrichment of the Galaxy from first principles is challenging. Chemical enrichment is the outcome of multiple physical process occurring simultaneously over the lifetime of the Galaxy, and most of these processes are poorly understood.
Some aspects of modelling the chemical evolution like the distribution of stellar masses and the fraction of stars dying at any given time are dictated by the initial mass function and the theory of stellar evolution, which are known relatively well \citep{2014PhR...539...49K, 2017ApJ...835...77M}. However, some of the other crucial aspects are not, including the star formation history, the radial distribution of gas, the in-fall of fresh gas, the outflow of enriched gas, the radial transport of enriched gas in the disc and the nucleosynthetic yields of different elements.

The limitations of chemical evolution models has prompted the use of data driven/empirical methods to disentangle the information hidden in the chemical abundances of Galactic stars. Examples of data driven approaches that exclusively explore the elemental abundance space are \citet{2015ApJ...807..104T} and \citet{2019ApJ...887...73C}.  While these are useful for understanding the role of different nucleosynthetic processes, they are less useful for understanding the evolution as they do not directly take the age or the location of stars into account.

Empirical studies of how elemental abundances depend on age have also been performed, providing useful insights into the chemical evolution in the Galaxy.  Using a sample of 189 nearby dwarfs, \citet{1993A&A...275..101E} showed that for a given age and orbital radius (equivalent to metallicity because age and radius correlated for their sample), stars have a small dispersion (0.05 dex) in $[\alpha/{\rm Fe}]$.  They suggested that this indicates that the products of nuclear synthesis from supernovae are well mixed in the inter stellar medium (hereafter ISM). The scatter in [X/Fe] for heavy elements like Ba, Y, Zr, and Nd, for a given age and metallicity was also claimed to be compatible with measurement errors.
\citet{2015A&A...579A..52N}, using 21 solar twins with better measurement precision and more elements, reported a small scatter in elemental abundances for a given age \citep[see also][]{2012A&A...542A..84D, 2017A&A...608A.112N}. $s$-process element Y was also shown to have tight correlation with age. \citet{2018ApJ...865...68B} further expanded the analysis of solar twins using 79 stars and studying trends of 30 elements. They showed that stars having similar age and metallicity have nearly identical elemental abundances \citep[see also][]{2016A&A...593A.125S,2018MNRAS.474.2580S}.  The high precision analysis of age-abundance relations was extended to stars with a wide range in metallicities by several studies \citet{2017MNRAS.465L.109F,2019A&A...624A..78D, 2020A&A...639A.127C}. They showed that a number of elements show abundance trends with age and much of the scatter for a given age can be attributed to metallicity. \citet{2020MNRAS.491.2043L}, using data from GALAH DR2, also explored trends with both age and metallicity.

In a significant development, \citet{2019ApJ...883..177N} showed that abundances of 17 elements can be predicted from just age and metallicity, with intrinsic scatter about the predictions being only about 0.02 dex. Importantly, they showed that the abundance-age-metallicity relations are valid well beyond the solar Galactocentric radius. Unlike previous studies, which were based on dwarfs and main-sequence turn-off (MSTO) stars, they used red clump stars and hence they also verified that the abundance relations are not exclusive to dwarfs.
However, \citet{2019ApJ...883..177N} only studied the low-$\alpha$ population, restricted to a narrow range in age and $\alpha$ abundances.  They studied mainly $\alpha$ and iron peak elements, which are thought to be produced by SNe II and SNe Ia explosions. Elements produced by other mechanisms, for example $s$-process elements from asymptotic giant branch (AGB) stars or $r$-process elements were not studied.  Finally, in their study age measurements that were independent of abundances, were only available for stars in the solar Galactocentric radius.

The GALAH+ survey, which has measured abundances of 30 elements for about 700,000 stars, has more than 50,000 MSTO stars and 4000 asteroseismic giants with reliable age and abundance measurements. This provides a new opportunity to study the dependence of elemental abundances on age and metallicity and to overcome the limitations of the previous studies, which is the aim of this paper.  Using this data
we expand the study with significantly more stars than previous studies
and to regions beyond the solar annulus.  Compared to \citet{2019ApJ...883..177N}, we also expand the analysis to include $s$ and $r$ process elements. Additionally, we compare and contrast the abundance trends for dwarfs with that for giants, whose ages are measured using independent techniques.

\begin{figure}
\centering \includegraphics[width=0.49\textwidth]{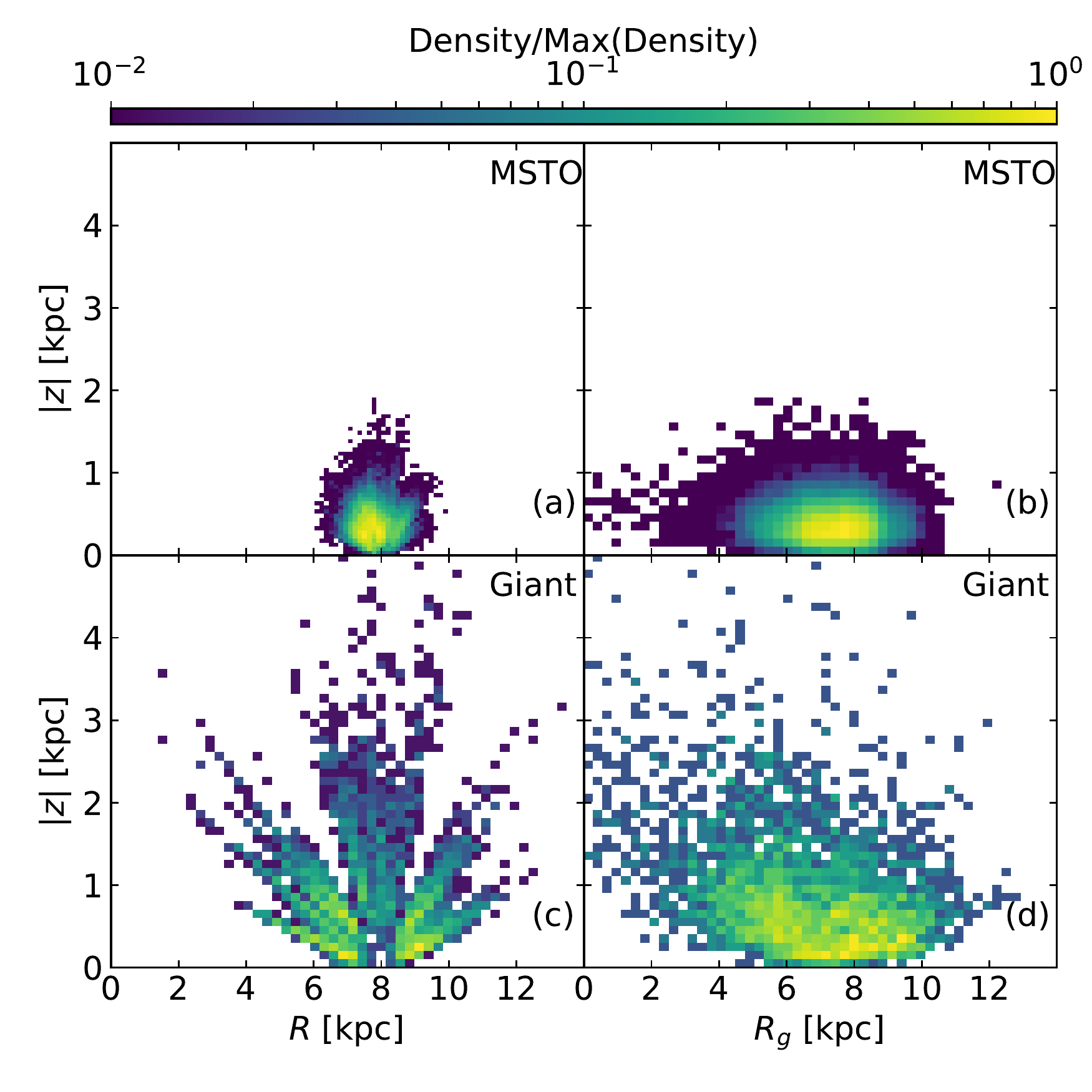}
\caption{Distribution of MSTO (top panels) and giant stars (bottom panels) analyzed in this paper shown in the $(R,|z|)$ (left) and $(R_g,|z|)$ (right) planes. The stars have SNR$>20$.
\label{fig:Rz_dist}}
\end{figure}

\section{Data}
In this paper we use data from the GALAH+ DR3 catalog, \citep{2020arXiv201102505B}, which in addition to the main GALAH survey also includes data from the TESS-HERMES \citep{2018MNRAS.473.2004S} and K2-HERMES \citep{2019MNRAS.490.5335S} surveys, which use the same spectrograph, observational setup, and data reduction pipeline as the GALAH survey. The catalog provides spectroscopic stellar parameters \teff, $\log g$, and [Fe/H] and abundances for 30 elements.  We focus on two specific type of stars, MSTO stars and asteroseismic red giants (RG) for which ages can reliably determined.  The RG stars have asteroseismic information from the NASA K2 mission and their spectroscopic follow-up was carried out by the K2-HERMES survey.

The ages and distances for the MSTO and RG
stars are computed with the BSTEP code \citep{2018MNRAS.473.2004S}. BSTEP provides a Bayesian estimate of intrinsic stellar parameters from observed parameters by making use of stellar isochrones.  For results presented in this paper, we use the PARSEC-COLIBRI stellar isochrones \citep{2017ApJ...835...77M}.  For the MSTO stars, we use the following observables, \teff, $\log g, [{\rm Fe/H}], [\alpha/{\rm Fe}]$, $J$, $Ks$, and parallax. For the RG stars, in addition to the above observables, we use the asteroseismic observables \dnu\ and \numax. These stars were observed
as part of the K2 Galactic Archaeology Program \citep{2015ApJ...809L...3S} and includes stars from campaigns 1 to 15 \citep{2017ApJ...835...83S}.
The asteroseismic analysis is conducted with the method by \citet{2010A&A...522A...1K,2014A&A...570A..41K}, known as the CAN pipeline. \dnu\ and \numax\ for the model stars along the isochrones are determined with the ASFGRID code \citet{2016ApJ...822...15S} that incorporates corrections to the \dnu\ scaling relation suggested by stellar models.

The RG stars were selected using the following selection function
\be
(1 < \log g < 3.5)\&(3500 < {\rm T}_{\rm eff}/{\rm K} < 5500)\&(SNR>10).
\ee
The MSTO stars were selected using the following selection
function
\be
(3.2 < \log g < 4.1)\&(5000 < {\rm T}_{\rm eff}/{\rm K} < 6100)\&(SNR>10)
\ee
to ensure they have reliable ages and chemical abundances.  The (age,[Fe/H]) plane is not populated uniformly, with a peak density around solar age and metallicity and the density falls off rapidly away from the peak.  Given non-negligible uncertainty on both age (~14\% )and metallicity (~0.08 dex), stars in the low-density boundary regions are most likely to have wrong age and metallicity and are not useful for studying abundance trends. To exclude stars in low density regions, we binned the stars in the (age,[Fe/H]) plane in the range 0 to 14 Gyr and -0.8 to 0.5 dex with 40 bins along each dimension and stars lying in bins having less than 40 stars were excluded. For RG stars, no such culling was performed due to the small size of this sample. For most of our analysis, we further restrict our sample to SNR$>20$, and with this restriction we had 50019 MSTO stars and 3708 RG stars.

In order to test if the abundance-age-metallicity relations are local or valid over the Galaxy, it is important to sample different Galactocentric $(R,z)$ locations. The distribution in the Galactocentric $(R,z)$ plane of stars that we use in this paper is shown in \autoref{fig:Rz_dist}. Also shown is the distribution in the $(R_g,z)$ plane, where $R_g=R (v_{\rm rot}/232.0 {\rm km/s})$ is the guiding radius and  $v_{\rm rot}$ is the azimuthal rotational velocity.  The MSTO stars are intrinsically faint and hence are mainly confined within 1 kpc from the Sun.  However, they do span a wide range in guiding radius, which makes it possible to study stars that were born beyond the solar neighborhood.  The RG stars span a wide range in both $R$ and $z$ and their $R_g$-span is also much larger.

\begin{figure*}
\centering \includegraphics[width=0.99\textwidth]{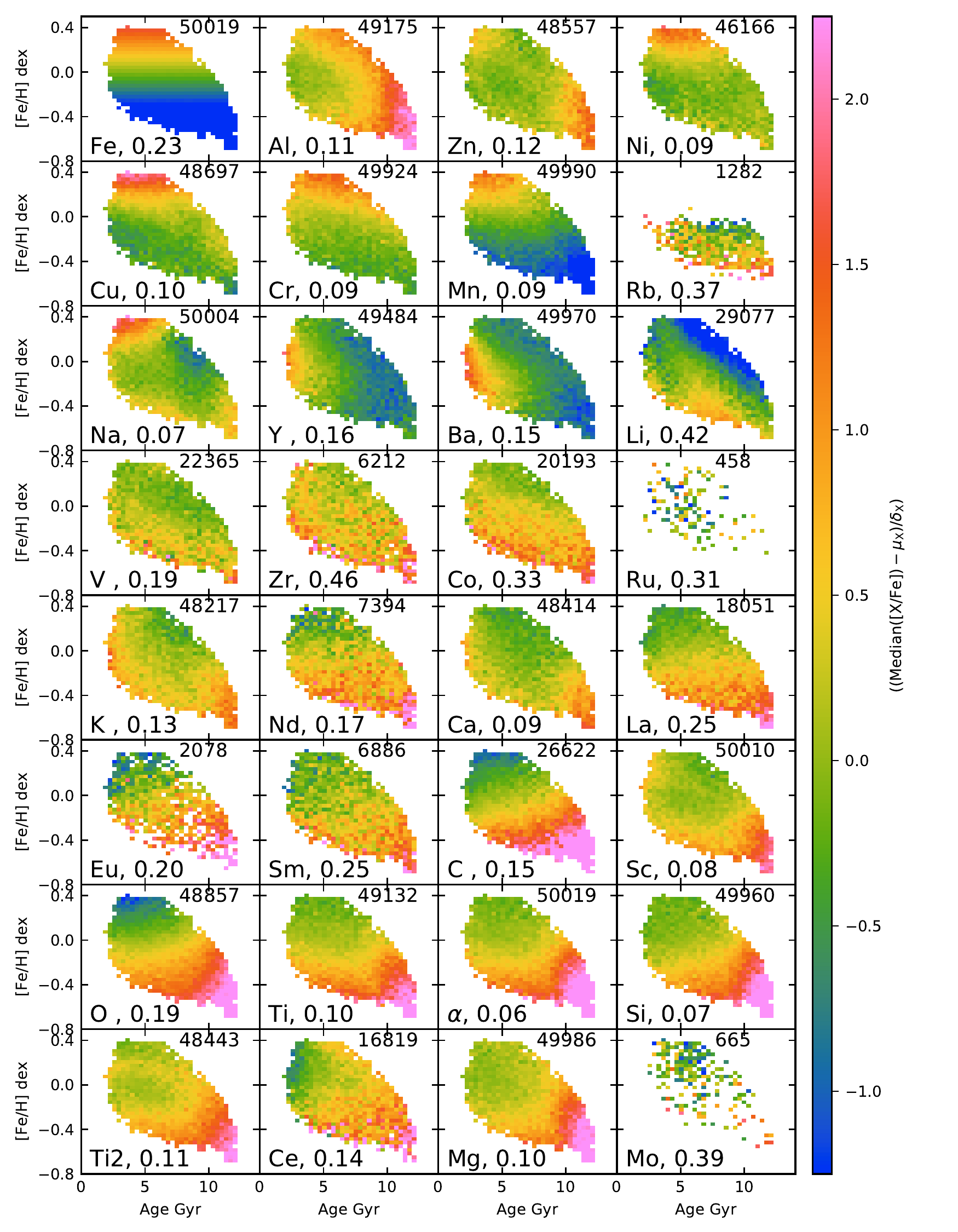}
\caption{Median abundance of different elements in the (Age,[Fe/H]) plane for MSTO stars with SNR$>20$. We standardize the median abundance in each bin by subtracting $\mu_{\rm X}$, the 50 percentile value, and then dividing by $\delta_{\rm X}$, the dispersion based on 16 and 84 percentile values. In each panel, the number of stars plotted is shown in the upper right and the $\delta_{\rm X}$ is shown in the lower right.
\label{fig:age_feh_map_abund_median_msto}}
\end{figure*}

\begin{figure*}
\centering \includegraphics[width=0.99\textwidth]{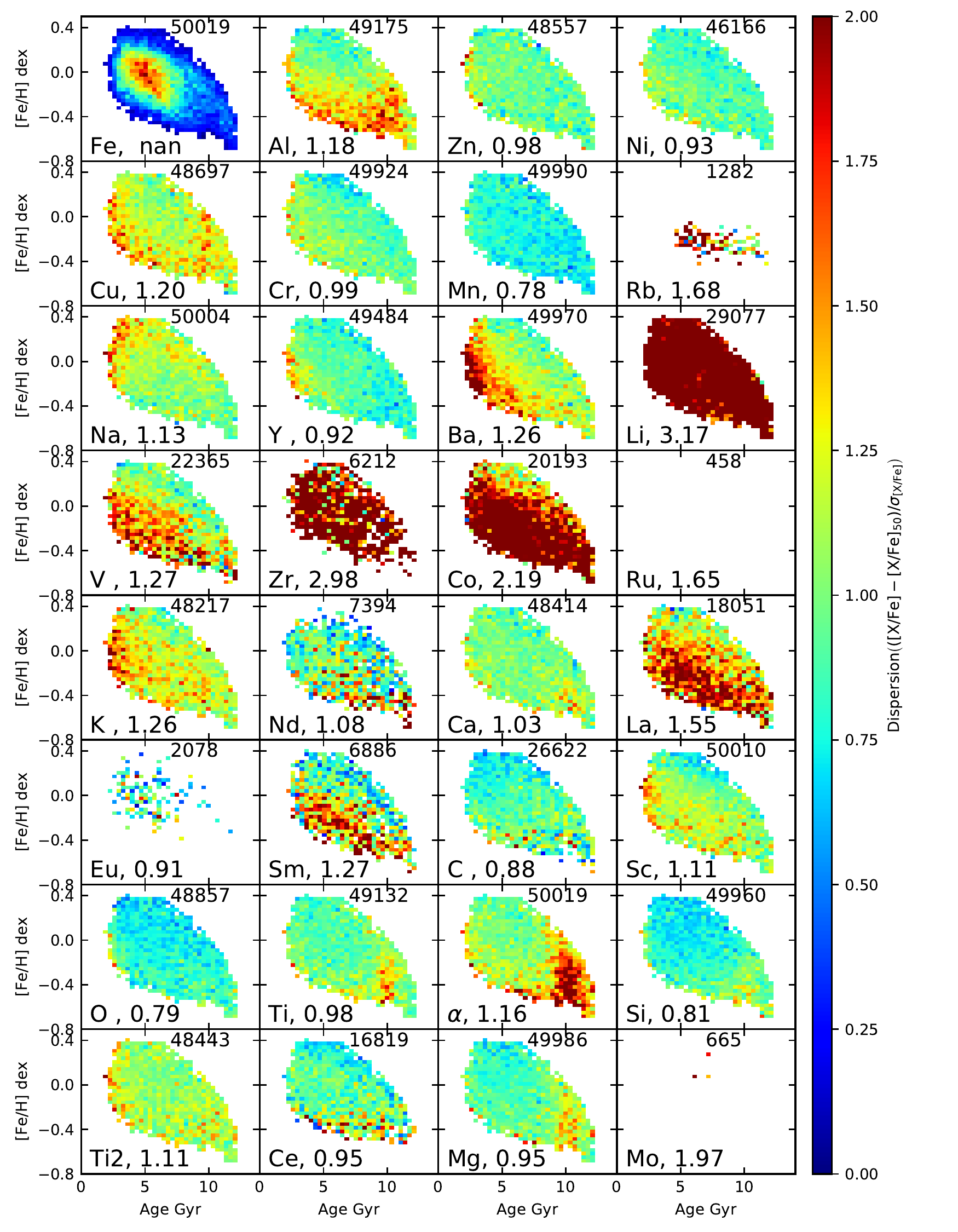}
\caption{Dispersion of elemental abundances  in the (Age,[Fe/H]) plane for MSTO stars with SNR$>20$. The dispersion shown is around median values and is normalized by dividing with the uncertainty of each data point.  The first panel shows the density of stars in (Age,[Fe/H]) plane. In each panel, the number of stars plotted is shown in the upper right and the average dispersion relative to uncertainty for all stars is shown in the lower right.
\label{fig:age_feh_map_abund_std_msto}}
\end{figure*}

\begin{figure*}
\centering \includegraphics[width=0.99\textwidth]{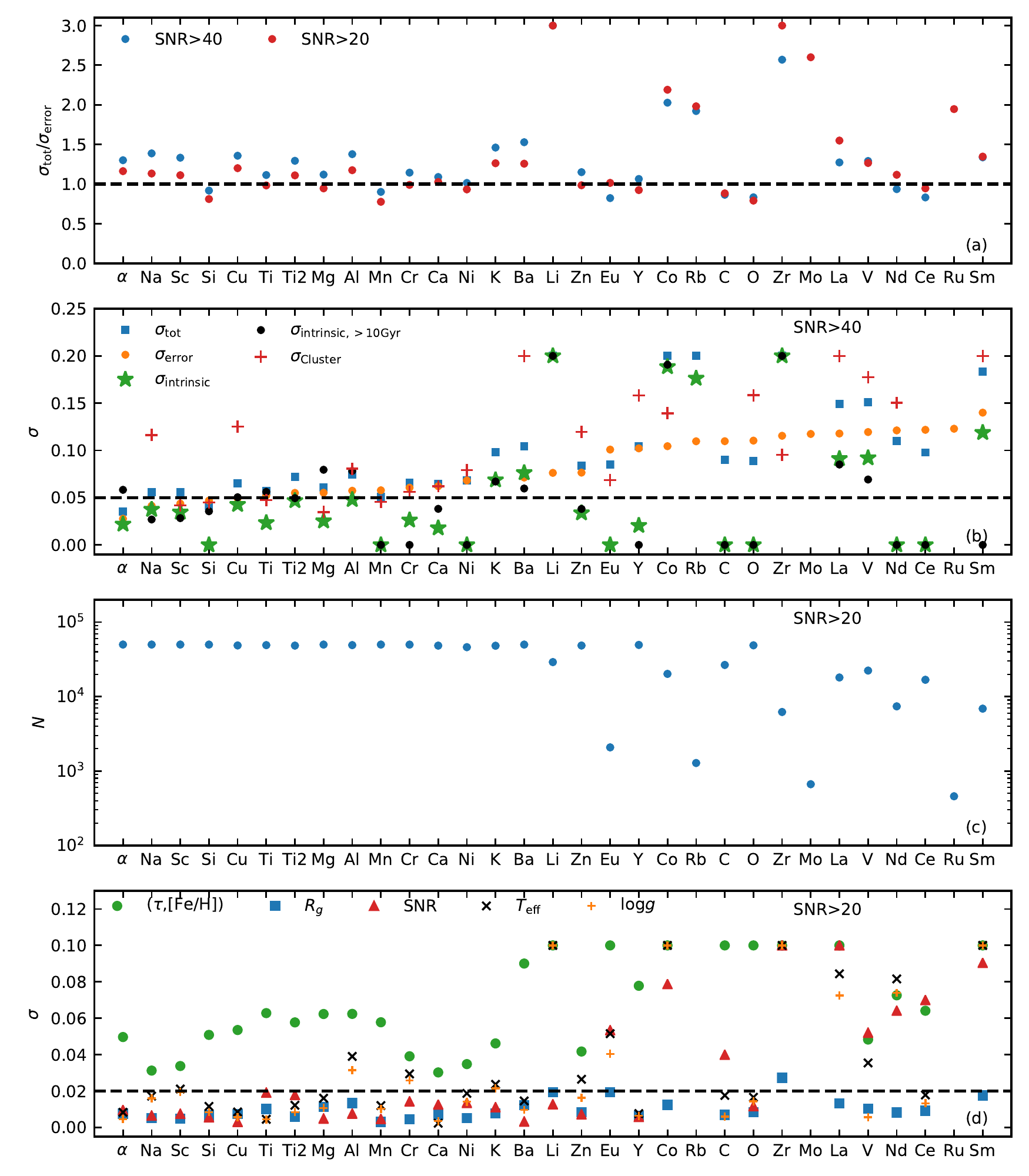}
\caption{Dispersion of elemental abundances for MSTO stars. (a) The dispersion about predicted values (based on age and [Fe/H]) relative to uncertainty for two signal to noise ratios.  (b) Dispersion of abundance about predicted values. Median uncertainty $\sigma_{\rm error}$ and an estimate of intrinsic dispersion $\sigma_{\rm intrinsic}=\sqrt(\sigma_{\rm tot}^2-\sigma_{\rm error}^2)$ is also shown alongside.  (c) Number of stars for which a measurement of the element exists. (d) Dispersion that can be attributed to a given variable. For stellar variables, e.g., guiding radius $R_g$, SNR and $T_{\rm eff}$ it is computed on residual abundance (observed abundance minus abundance predicted by age and metallicity)
\label{fig:age_feh_abund_stats_msto}}
\end{figure*}

\begin{figure*}
\centering \includegraphics[width=0.99\textwidth]{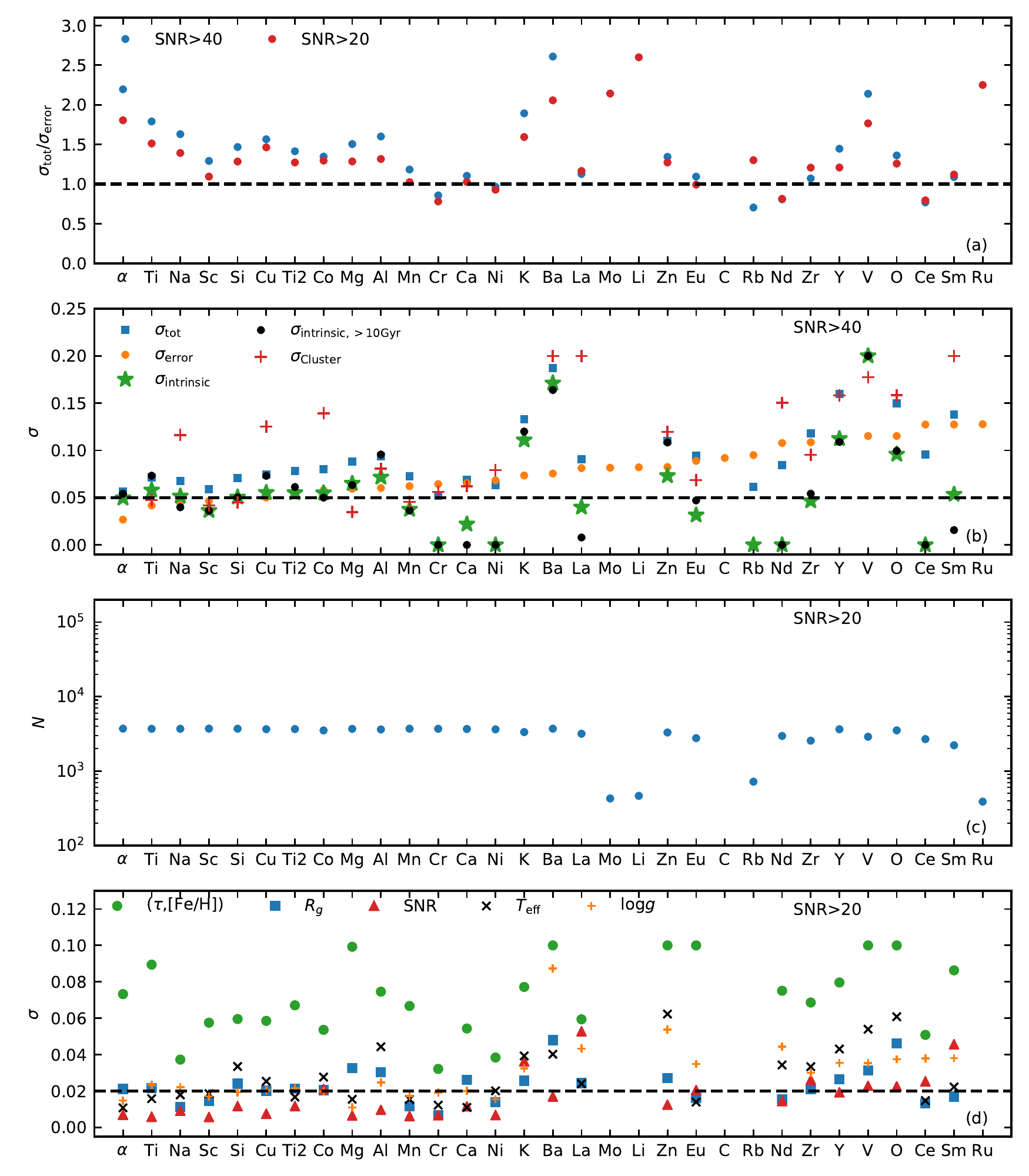}
\caption{Same as \autoref{fig:age_feh_abund_stats_msto} but for asteroseismic giants.
\label{fig:age_feh_abund_stats_giant}}
\end{figure*}

\section{Results}
\subsection{Abundance trends with age and [Fe/H]}
In \autoref{fig:age_feh_map_abund_median_msto} we show the median chemical abundances in the (age, [Fe/H]) plane for MSTO stars. The abundances are shifted to be zero for solar age and metallicity, and are scaled with respect to the global spread (see figure caption). A wide variety of trends can be seen.  Some pronounced examples are: (1) Elements Ba, Y, and Al show a gradient along the horizontal age axis, indicative of strong dependence on age and weak dependence on [Fe/H]. (2) Elements Cu, Mn, Cr, and Ni
show a gradient along the vertical [Fe/H] axis, indicative of strong dependence on [Fe/H] and weak dependence on age. (3) Elements Mg, Si, Ca, Ti, O, and Sc show a diagonal gradient from top left to bottom right, indicative of strong dependence on both age and [Fe/H]. 4) Some have positive gradient
with age (Ba and Y) while others have negative (Mg, Si). 5) Dependence of some elements is more complicated, for example Na has a gradient with [Fe/H] that is positive for young stars and negative for older stars.

\subsection{Scatter of abundances at fixed age and [Fe/H]}
In \autoref{fig:age_feh_map_abund_std_msto}, we show the abundance dispersion in the (Age, [Fe/H]) plane for MSTO stars. The dispersion is shown relative to measurement uncertainty. The relative dispersion computed over all stars is labelled in each panel. All elements, except for Co, La, Li, and Zr; have relative dispersion that is small (less than 1.3). This suggests that  the abundances can be predicted from age and metallicity.  Very high dispersion for Co, Li, and Zr suggests there are other processes in addition to Galactic chemical evolution that play important roles in the chemical finger print we observe. For example, the abundance of Li is known to depend on surface gravity and temperature \citep{2020MNRAS.497L..30G}.

For most elements, the dispersion is close to 1 over most regions of the (Age, [Fe/H]) plane. However, some elements show slightly higher dispersion (dark brown regions) in certain specific regions. One reason for this could be underestimated uncertainties, which can spuriously increase the dispersion, as we compute it relative to uncertainty. However, the uncertainty was found to be same over the whole plane, and hence this cause can be ruled out.  A more probable reason for higher dispersion at a given age and metallicity is contamination from stars with incorrect age and/or metallicity. The top left most panel of \autoref{fig:age_feh_map_abund_std_msto} shows that the distribution of stars in the (age, [Fe/H]) plane is highly non uniform. A bin in an under dense region can have significant contamination from stars having incorrect age and/or metallicity determination (meaning different from that corresponding to the bin). This can be seen very clearly for Cu, where the dark region resembles a low density contour from the first panel of \autoref{fig:age_feh_map_abund_std_msto}. Wherever the dependence on either age or [Fe/H] is steep, one can expect higher dispersion in the low density regions.  Co, La, V, and Sm all have strong dependence on [Fe/H] at the metal poor end and show higher dispersion towards the low metallicity edge.  Mg, Si, Ti, Ca, Cu, and Al all show a sharp rise with age at around 10 Gyr, a region also having low density of stars and they all show higher dispersion at around this age.  Ba and Y show a strong dependence on age for ages less than 4 Gyr and they have a higher dispersion at the low age edge. For some elements, for example Ba, the large dispersion for young stars could also be related to chromospheric activity which can significantly alter the abundance of an element \citep{2019MNRAS.490L..86Y, 2020ApJ...895...52S}.

\begin{figure*}
\centering \includegraphics[width=0.99\textwidth]{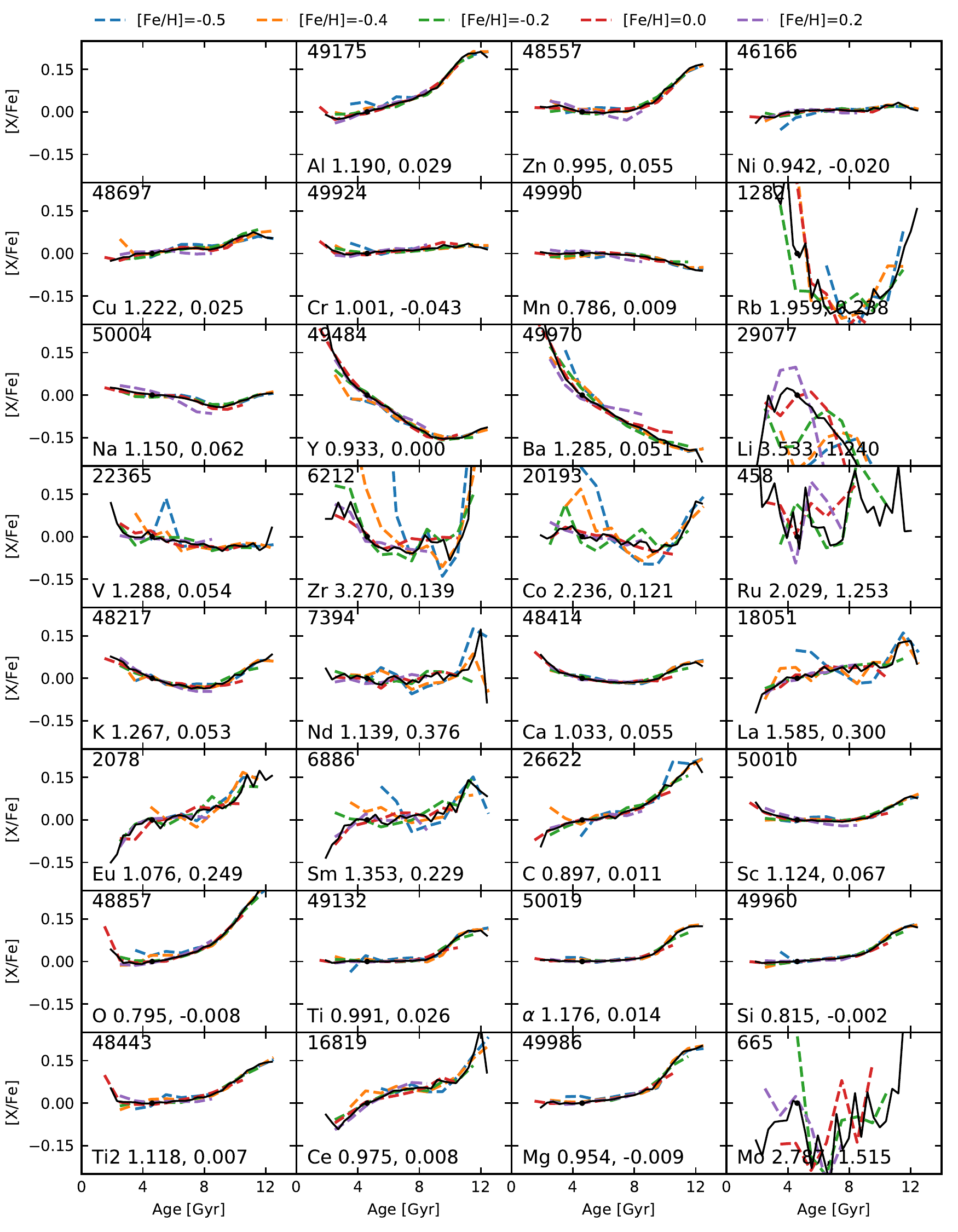}
\caption{ Elemental abundance $f_{\tau,{\rm X}}$ of MSTO stars (SNR>10) as a function of age for different elements (solid black line).  Dashed curves show the dependence with age for different bins in metallicity. A metallicity dependent shift $f_{\rm [Fe/H],X}$ is applied to the dashed curves.  The number on the top denotes the number of stars. The first number on the bottom denotes the overall dispersion (based on 16 and 84 percentile values) of the measured abundance as compared to abundance predicted by the model. The profiles are constrained to have zero abundance for stars with solar age and metallicity.  The abundance for solar age and metallicity is given by the second number on the bottom.
\label{fig:age_abund_relation_msto}}
\end{figure*}

In \autoref{fig:age_feh_abund_stats_msto}a we show the dispersion of abundances around the predicted values ($\sigma_{\rm tot}$) relative to measurement uncertainty ($\sigma_{\rm error}$) for MSTO stars. The elements are sorted by $\sigma_{\rm error}$. To account for the fact that the uncertainty varies from star to star, we estimate the dispersion of ${\rm([X/Fe]-[X/Fe]_{pred}}/\sigma_{\rm error})$,
which corresponds to $\sigma_{\rm tot}/\sigma_{\rm error}$ when $\sigma_{\rm error}$ is a constant.
Here, ${\rm [X/Fe]_{pred}}$ is the value predicted from median abundance maps in the (age,[Fe/H]) plane (see \autoref{fig:age_abund_relation_msto}).
The relative dispersion $\sigma_{\rm tot}/\sigma_{\rm error}$ was found to be close to 1, suggesting that the total dispersion is similar to that of the measurement uncertainty.  A slight dependence on SNR (green channel) can be seen as elements having ratio less than 1 for low SNR, suggesting that we might be overestimating the uncertainty at low SNR. For this reason, in \autoref{fig:age_feh_abund_stats_msto}b, where  we analyse absolute dispersion, we restrict our analysis to stars with SNR>40.

In \autoref{fig:age_feh_abund_stats_msto}b we show $\sigma_{\rm tot}$ alongside the median measurement uncertainty $\sigma_{\rm error}$ and intrinsic the dispersion  $\sigma_{\rm intrinsic}=\sqrt{\sigma_{\rm tot}^2-\sigma_{\rm error}^2}$ for all MSTO stars as well as a subset of them with age greater than 10 Gyr.  The total dispersion is similar to that of the measurement uncertainty. The intrinsic dispersion (green points) is below 0.05 dex for most elements.  K, Ba, La, and V have mildly higher dispersions (between 0.05 and 0.1) while Li, Co, and Zr have very high dispersion (greater than 0.15). Overall 23 elements had $\sigma_{\rm intrinsic}$ less than 0.12 with a mean of 0.033. The $\sigma_{\rm intrinsic}$ for stars older than 10 Gyr is also similar to that of all stars, but for $\alpha$ elements it was systematically higher, around 0.05 dex instead of 0.02 dex. This is most likely due the combination of the following two factors for the old stars, the rate of change of [$\alpha$/Fe] with age being very high (see \autoref{fig:age_abund_relation_msto}) and the uncertainty on age being non negligible. This is supported by the fact that for other other elements (e.g. Ba, Y, Na, K) which do not vary sharply with age at old ages, there was either a reduction in $\sigma_{\rm intrinsic}$ or no change at all.  The dispersion in open clusters observed by GALAH (red plus points) is also similar to $\sigma_{\rm tot}$. Some elements show slightly larger dispersion for the open cluster data. This could be because the open cluster data contains both MSTO and giants, while we here focus on only the MSTO stars. Also, this could be
because the open cluster stars are in general young, and young stars rotate faster and they have strong chromospheric activity  that makes the abundance analysis uncertain.
For some elements, like Rb, Mo and Ru, the intrinsic scatter is difficult to study, as there are very few stars with their measurements (< 1500). This indicates that these
elements are difficult to measure with our setup. In general such elements also have larger
uncertainty. This is seen in \autoref{fig:age_feh_abund_stats_msto}c, where we show the number of stars for which the abundance of an element has been measured. The number of stars with measured elemental abundances decreases with increasing $\sigma_{\rm error}$.

If the abundance of an element is a function of age and metallicity only, then the abundance should have no dependence on any other variable. However, we do find some dependence on variables like guiding radius $R_g$, SNR, \teff, and $\log g$. This is shown in \autoref{fig:age_feh_abund_stats_msto}d, where we plot the dispersion that can be attributed to each of the above variables.  The dispersion that can attributed jointly to age and [Fe/H] is also shown. This dispersion measures the abundance information associated with an independent variable or jointly with two independent variables. Suppose $y$ is a variable that depends on a variable $x$, or more generally a vector ${\bf x}$ for more than one variables, and the relationship can be described by a function $y({\bf x})$.  The dispersion attributable to ${\bf x}$ is given by $\sqrt{<y({\bf x})^2>}$. To derive $y({\bf x})$, we measure the median abundance in bins of a given variable $x$ (mutidimensional binning for ${\bf x}$) and then use interpolation to predict the relationship. For variables other than age and metallicity, we compute the dispersion of the residual abundance, meaning the observed abundance minus the abundance predicted by age and metallicity.

In \autoref{fig:age_feh_abund_stats_msto}d the dispersion due to joint dependence on age and metallicity (green dots) is the dominant one for most elements. Elements left of Li (having small $\sigma_{\rm error}$), have small dispersion (less than 0.02) in variables other than age and metallicity. There are some exceptions. A relatively strong dependence can be seen for Ti on SNR, and for Al, Cr, and K on \teff\ and $\log g$.  Elements to the right of Li (having large $\sigma_{\rm error}$), with the exception of Y and O, show large dispersion (greater than 0.05) in at least one of the following, SNR, \teff,
or $\log g$.

We now look at the dispersion results for giants, which is shown in \autoref{fig:age_feh_abund_stats_giant}.  Just like MSTO stars the ratio $\sigma_{\rm tot}/\sigma_{\rm error}$ is less than 1.5 and the $\sigma_{\rm intrinsic}$ is less than 0.08 for most elements. K, Ba, Y, and V show large dispersion (greater than 0.1). Mo, Li, C, Rb, and Ru are measured for less than 500 stars and we deem their measurements to be less reliable. Overall 23 elements had $\sigma_{\rm intrinsic}$ less than 0.12 with a mean of 0.046. In general, $\sigma_{\rm intrinsic}$ is slightly higher than that for MSTO stars.  In \autoref{fig:age_feh_abund_stats_msto}d the dependence on age and metallicity is clearly stronger than other variables. However, the  dependence on variables like $R_g$, SNR, \teff, and $\log g$ is stronger than that seen for MSTO stars. For elements left of La that are measured with good precision ($\sigma_{\rm error}$< 0.08), dependence on $R_g$, SNR, \teff, and $\log g$ is weak. However, a moderately strong dependence o \teff\ can be seen for Si, Al, and K.

To understand the origin of the small scatter in abundances, we estimated abundances by binning the stars in the ([$s$/Fe]-[$\alpha$/Fe], [Fe/H]) plane and interpolation over the median values in each bin, with $[s{\rm /Fe}]=({\rm [Ba/Fe]+[Y/Fe]})/2$.  The estimates of $\sigma_{\rm intrinsic}$ were very similar but systematically lower than when estimating from abundances from age and [Fe/H].  For MSTO stars, 23 elements had $\sigma_{\rm intrinsic}$ less than 0.12 with a mean of 0.026.  For giants stars 24 elements had $\sigma_{\rm intrinsic}$ less than 0.12 with a mean of 0.035. This suggests that there are only 3 independent groups of elements from which most of the abundance scatter can be explained; the Fe group, the $\alpha$-process group, and the $s$-process group.

\begin{figure*}
\centering \includegraphics[width=0.99\textwidth]{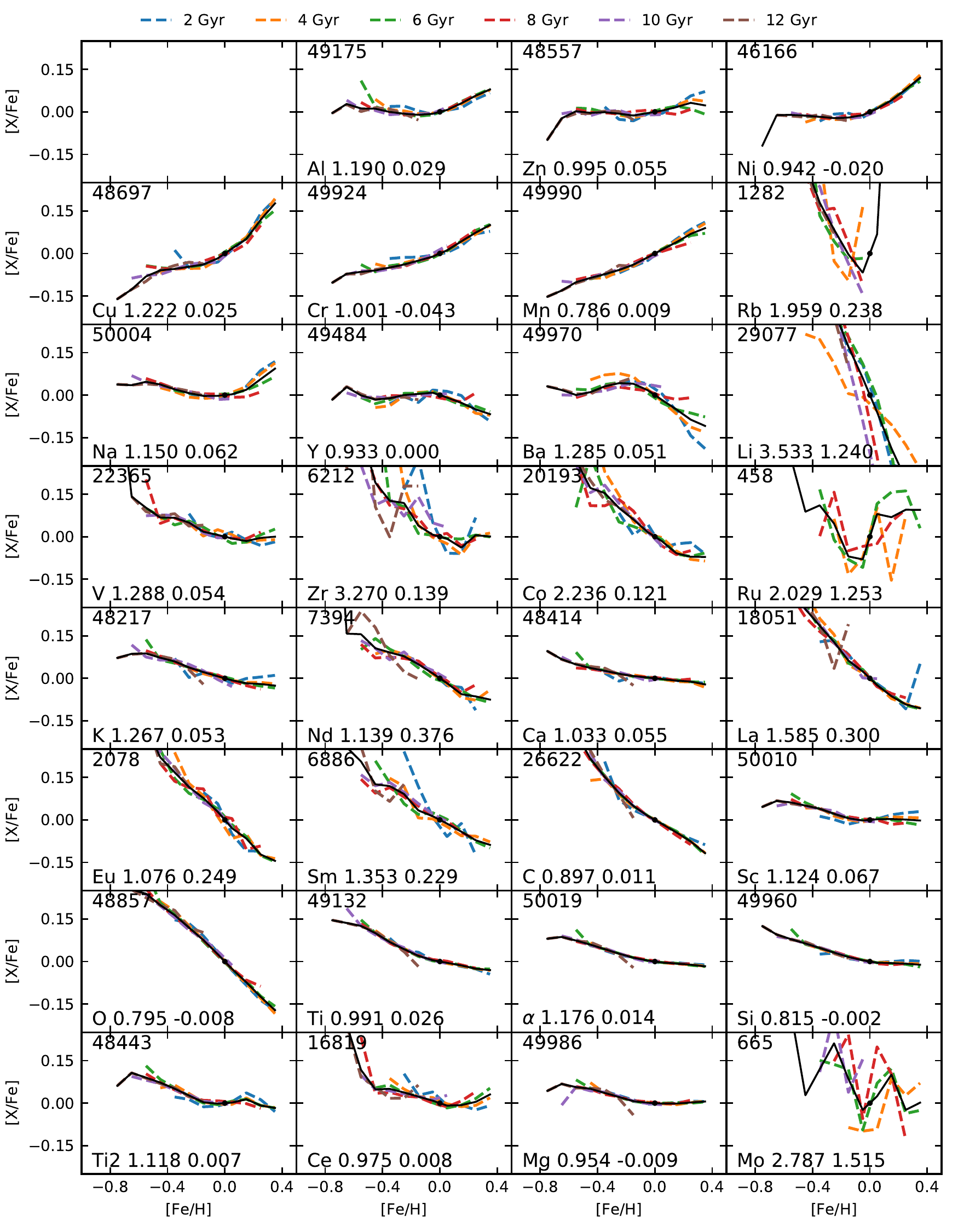}
\caption{Elemental abundance $f_{\rm [Fe/H], X}$ of MSTO stars (SNR>10) as a function of [Fe/H] for different elements (solid black line). Dashed curves show the dependence with [Fe/H] for different bins in age. An age dependent shift $f_{\tau,{\rm X}}$ is applied to the dashed curves.  The number on the top denotes the number of stars. The first number on the bottom denotes the overall dispersion (based on 16 and 84 percentile values) of the measured abundance as compared to abundance predicted by the model. The profiles are constrained to have zero abundance for stars with solar age and metallicity.  The abundance for solar age and metallicity is given by the second number on the bottom.
\label{fig:feh_abund_relation_msto}}
\end{figure*}

\subsection{An empirical model for abundance as a function of age and metallicity}
\label{sec:empirical}
From previous sections we have seen that elemental abundances can indeed be predicted from age and [Fe/H]. In this section we build an empirical model to describe the dependence of abundance on age and metallicity. The most flexible approach is to bin the stars in age and metallicity compute the median abundance in each bin and then use 2d interpolation in the (age,[Fe/H]) plane. This is fully non-parametric but has too many degrees of freedom.  We adopt a slightly less flexible approach, which has fewer degrees of freedom, but was found to be equally predictive. The abundance [X/Fe] of an element is postulated to be an additively separable function of age, $\tau$, and metallicity, [Fe/H],
\be
{\rm [X/Fe]}  =  f_{\odot,{\rm X}}+f_{\rm [Fe/H],{\rm X}}({\rm [Fe/H]})+f_{\tau,{\rm X}}(\tau).
\label{equ:empirical_model}
\ee
Here $f_{\odot,{\rm X}}$ is a constant specifying the average elemental abundance of disc stars of solar age and metallicity. The functions on the right hand side, being additive, are degenerate with respect to a constant. Hence, they are forced to satisfy
\be
f_{\tau,{\rm X}}(\tau_{\odot})=f_{\rm [Fe/H],X}(0.0)=0, \textrm{with $\tau_{\odot}=4.6$ Gyr}
\ee

If the Sun is a typical disc star, $f_{\odot,{\rm X}}$ should be zero because abundances are defined relative to the solar abundance.  However, if the measured abundances have a dependence on stellar parameters like \teff\ and $\log g$, then $f_{\odot,{\rm X}}$ will vary depending upon the mean \teff\ and $\log g$ of stars with solar age and metallicity in the studied sample.
Such a dependence on stellar parameters could be due to systematics in spectroscopic analysis or due to real physical effects, such as atomic diffusion \citep{2017ApJ...840...99D,2019A&A...627A.117L}. Alternatively, the Sun might be atypical for certain elements due to planet formation. For example, terrestrial planets are supposed to lock up refractory elements \citep{2009ApJ...704L..66M,2018MNRAS.473.2004S}.

To determine $f_{\rm [Fe/H],X}$ and $f_{\tau,{\rm X}}$, we adopt a non-parametric approach. The function to be determined is computed at predefined equispaced locations and linear interpolation is used to compute the values for any arbitrary location.  We start with guess values for $f_{\rm [Fe/H],X}$ (based on stars with age close to solar age). Next, we alternately estimate $f_{\tau,{\rm X}}$ and $f_{\rm [Fe/H],X}$ by computing median values of ${\rm [X/Fe]}  -  f_{\odot,{\rm X}}-f_{\rm [Fe/H],{\rm X}}({\rm [Fe/H]})$ in bins of $\tau$, and  median values of ${\rm [X/Fe]}  -  f_{\odot,{\rm X}} - f_{\tau,{\rm X}}(\tau)$ in bins of [Fe/H], respectively.

The estimated functions $f_{\tau,{\rm X}}$ and $f_{\rm [Fe/H], X}$ are shown as black lines in \autoref{fig:age_abund_relation_msto} and \autoref{fig:feh_abund_relation_msto}, respectively.  Also shown as dashed curves in \autoref{fig:age_abund_relation_msto} is the median abundance as a function of age for stars lying in different metallicity bins but shifted by subtracting $f_{\odot,{\rm X}}+f_{\rm [Fe/H], X}$. The curves for different metallicity bins closely follow the black curve corresponding to $f_{\tau,{\rm X}}$ suggesting that the dependence on age is independent of the dependence on metallicity. Similar conclusion can be reached from \autoref{fig:feh_abund_relation_msto} where the dashed curves, showing the abundance dependence on [Fe/H] for stars in different age bins but shifted by subtracting $f_{\odot,{\rm X}}+f_{\tau,{\rm X}}$, follow the solid black curve corresponding to $f_{\rm [Fe/H], X}$.

Stars with very low [Fe/H] (less than -0.6 dex) are also the oldest stars (age greater than 11.5 Gyr), see \autoref{fig:age_feh_map_abund_median_msto}a).  This means that $f_{\rm [Fe/H],X}$ is degenerate with $f_{\tau,{\rm X}}$ for the most metal poor and the oldest stars. This in turn means that $f_{\tau,{\rm X}}$ beyond 11.5 Gyr and $f_{\rm [Fe/H],X}$ below -0.6, could be inaccurate. However, their sum will still be accurate.

\begin{figure}
\centering \includegraphics[width=0.49\textwidth]{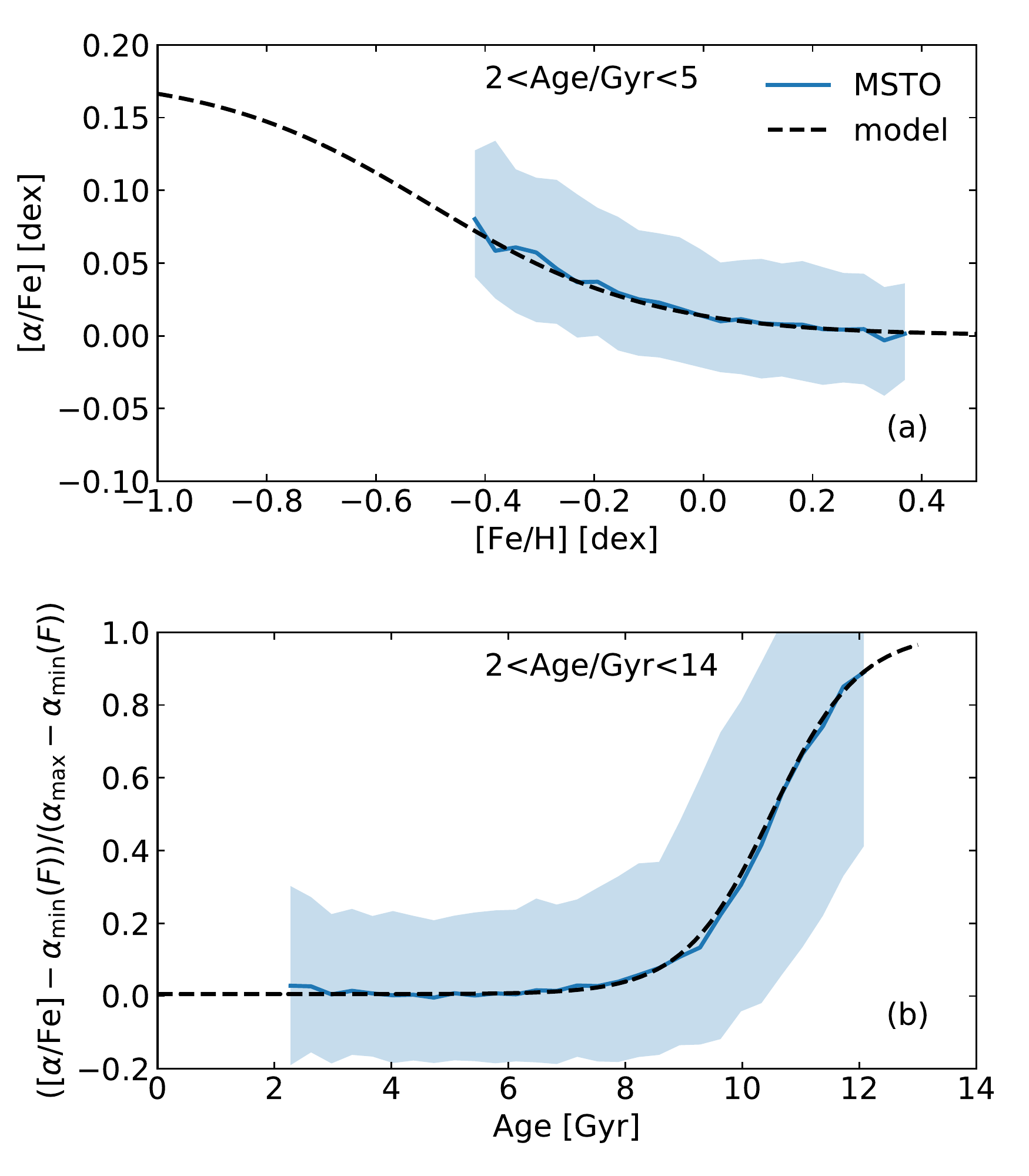}
\caption{ Metallicity and age dependence of $\alpha$ elemental abundance.
Solid blue line shows the median profile for the MSTO stars observed by GALAH and having SNR$>20$, and the blue shaded region is the 16 and 84 percentile spread around it. The dashed lines are predictions of a model with, $\alpha_{\rm max}=0.225$, $\alpha_{\rm outer}=0.18$, $F_{\alpha}=-0.5$, $\Delta F_{\alpha}=0.4$, $\tau_{\alpha}=10.5$ Gyr and $\Delta \tau_{\alpha}=1.5$, which is designed to fit the GALAH data. $F$ stands for metallicity [Fe/H], and $\alpha_{\rm min}(F)$ is an analytical function of metallicity as shown by the dashed line in panel (a).
\label{fig:alpha_feh_profile}}
\end{figure}

\subsubsection{An analytical model for [$\alpha$/Fe]}
In addition to the empirical model described by \autoref{equ:empirical_model} we
explored an additional model specifically for [$\alpha$/Fe]. As compared to previous model, this
model was analytic and has even fewer degrees of freedom. The model is motivated by the physics
of how the supernovae enrich the ISM.
From \autoref{fig:age_abund_relation_msto}, we see that [$\alpha$/Fe] is approximately constant with age till about 8 Gyr but rises rapidly thereafter. We postulate a $\tanh$ function that transitions from a low value $\alpha_{\rm min}$ to a high value $\alpha_{\rm max}$ at an age $t_{\alpha}$, with the sharpness of the transition being controlled by $\Delta t_{\alpha}$.
\be
[\alpha/{\rm Fe}](F,\tau)&=&\alpha_{\rm min}(F)+ \nonumber \\ && \frac{\alpha_{\rm max}-\alpha_{\rm min}(F)}{2}\left[{\rm tanh}\left(\frac{\tau-\tau_{\alpha}}{\Delta \tau_{\alpha}}\right)+1\right]
\ee
The relationship is shown as dashed line in \autoref{fig:alpha_feh_profile}b.
The relationship is motivated by the physics of chemical
enrichment (Fe and $\alpha$ elements) in the Galaxy which
is mainly regulated by Supernovaes. The initial value
$\alpha_{\rm max}$ of [$\alpha$/Fe]
is set by the yields of SNII, which occur almost
immediately (10 Myr) after the initiation of star formation at age $\tau_{\rm max}$. We expect $\alpha_{\rm max}$ to be independent
of metallicity $F$.
SNIa mostly produce Fe and almost no $\alpha$ elements,
which leads to a drop in [$\alpha$/Fe]. SNIa require a
binary companion and can only occur after significant
time delay. The SNIa rates typically peak about 1 Gyr
after star formation. This typically sets the time
scale $\Delta \tau_{\alpha}$ of transition from high
to low [$\alpha$/Fe]. We expect $\tau_{\alpha}$ to be given by
$\tau_{\rm max}-k\Delta \tau_{\alpha}$, with
$k$ being somewhere between 1 and 2, the exact value
needs to be determined by fitting to observational data.
As the evolution proceeds at some stage the ISM will
reach an equilibrium state due to infall of fresh metal poor gas
and this will set the floor $\alpha_{\rm min}$.
Since, the star formation rate and the infall rate
are not same at all birth radius, $\alpha_{\rm min}$ will
depend on birth radius. Given $F$ is a function of $R_b$
and $\tau$, we expect $\alpha_{\rm min}$ to be a function of $F$.

The functional form of the new analytic model is different from \autoref{equ:empirical_model}.
For age less than 8 Gyr both models are effectively similar. However for age greater than 8 Gyr,
their functional form differs but their predicted values for observed stars are
indistinguishable. This is because old stars are in general metal poor and this introduces
a degeneracy in the model parameters.
To really test the difference between
the two models we need old stars with a wide range of metallicity.

\begin{figure}
\centering \includegraphics[width=0.49\textwidth]{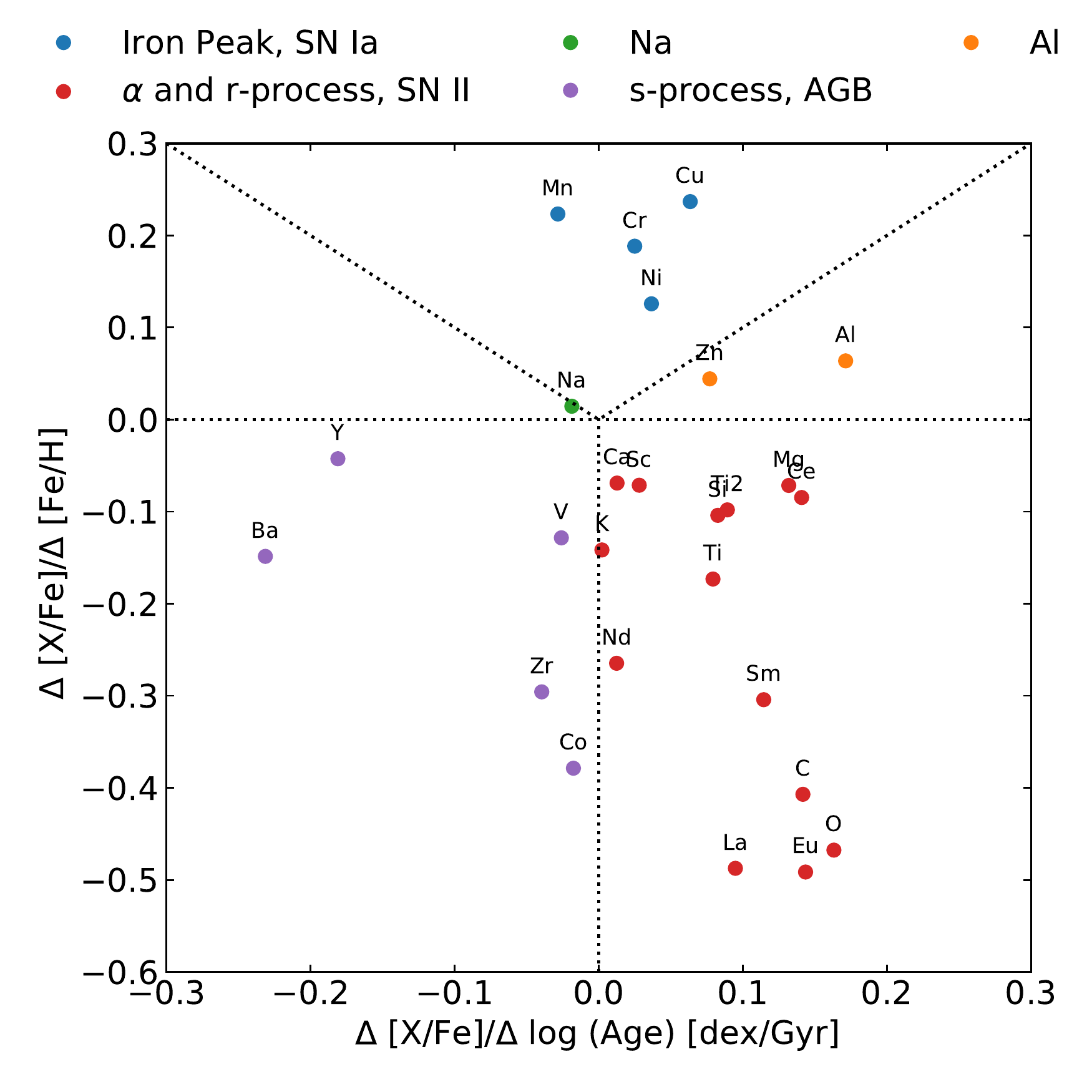}
\caption{Classification of elements based on the gradient of the abundance in the (age,[Fe/H]) plane. Probable production sites and nucleosynthetic processes are also listed.
\label{fig:classification}}
\end{figure}

\subsection{Classification of elements based on their abundance trends with age and [Fe/H].}
We now describe a simple scheme to classify the elements based on their trends with age and [Fe/H]. First, we compute $\Delta {\rm [X/Fe]}/\Delta \log {\rm Age}$, the rate of change of abundance with age, using $f_{\tau,{\rm X}}$ for an age of 3 and 11 Gyr.
Second, we compute $\Delta  {\rm [X/Fe]}/\Delta \log {\rm [Fe/H]}$, the rate of change of abundance with [Fe/H]), using $f_{\rm [Fe/H], X}$ for an [Fe/H]
of -0.4 and 0.2. Next, we make a scatter plot of different elements in the gradient plane, which is shown in \autoref{fig:classification}. The elements are classified based on the polar coordinate angle $\theta$ in the gradient plane. Al and Zn  ($\theta<45^{\circ}$) have positive age gradient but a weak [Fe/H] gradient. Ni, Cr, Cu, and Mn ($45^{\circ}<\theta<135^{\circ}$) have a strong [Fe/H] gradient. They are iron-peak elements and are produced by SNe Ia. The strong gradient with [Fe/H] suggest that the contribution of SNe Ia
to the production of these elements is more than that for Fe, which is about 50\% \citet{2020ApJ...900..179K}.
Na ($135^{\circ}<\theta<180^{\circ}$) has a mild negative age gradient. Ba, Y, V, K, Zr, and Co ($180^{\circ}<\theta<270^{\circ}$) have a negative age and [Fe/H] gradient. Among these elements Ba and Y are known to be elements produced by the $s$-process in AGB stars. As the contribution of AGB stars increases with time, the abundance of these elements is expected to have a negative age gradient. The last group of elements are the ones lying between $270^{\circ}<\theta<360^{\circ}$ and they are produced most likely by the $\alpha$ and the $r$-process in SNe II. Such elements are expected to have a positive age gradient. The s-process, $\alpha$-process, and  $r$-process elements are expected to have negative [Fe/H] gradients due to the following reason.  About 50\% of iron
is produced by SNe Ia, which is not the main source of the above elements, and since we measure the abundances relative iron, the abundances for the above elements will fall with increasing [Fe/H].

\begin{figure*}
\centering \includegraphics[width=0.99\textwidth]{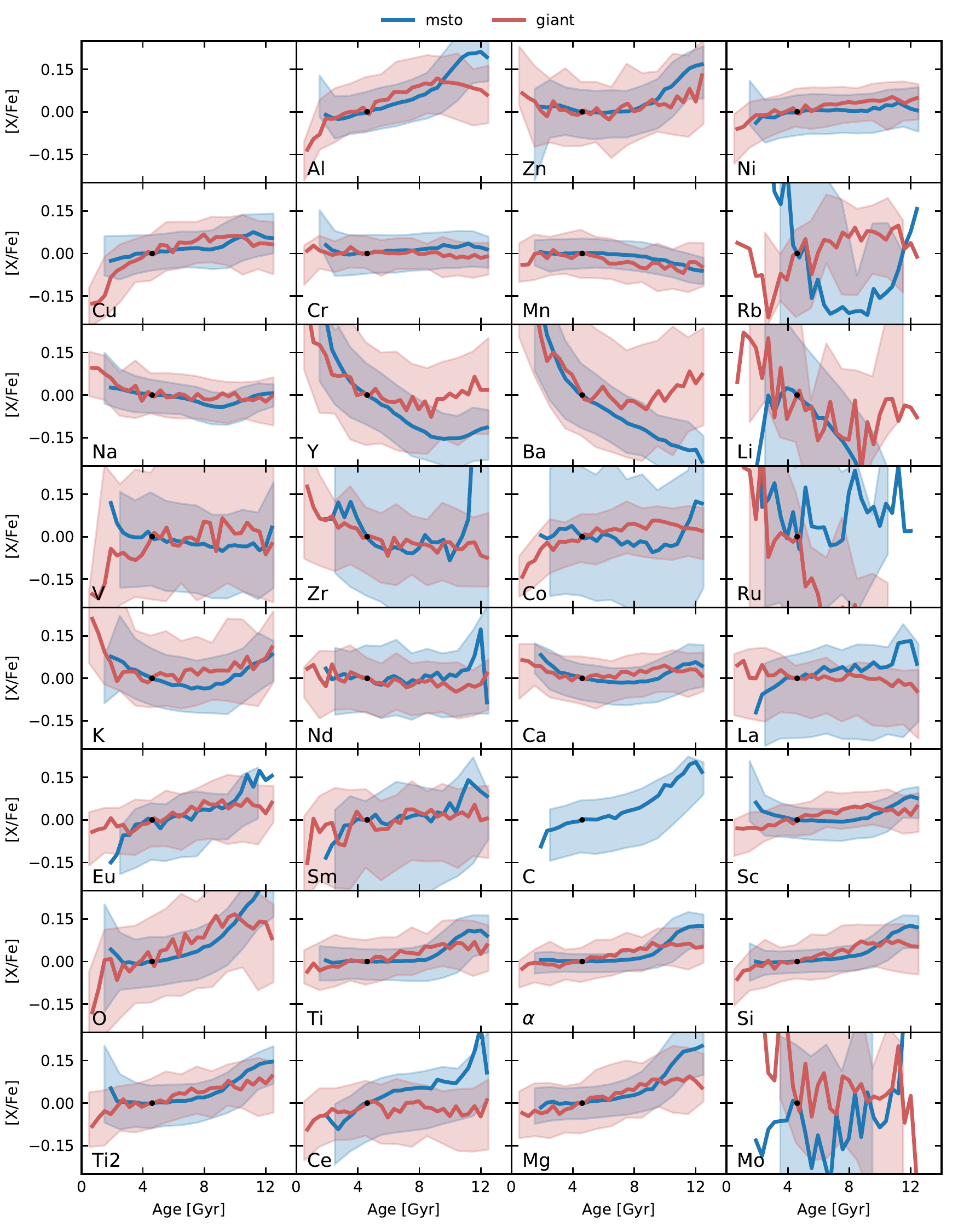}
\caption{Abundance of different elements as a function of age for MSTO (blue) and seismic giant (red) stars with SNR$>40$. The solid lines shows the function $f_{\tau,{\rm X}}$ as given by   \autoref{equ:empirical_model}.  The shaded region shows the 16 and 84 percentile spread around the relation (spread of [X/H]-$f_{\rm [Fe/H], X}-f_{\odot, {\rm X}}$) for stars with SNR$>40$.
The red profile is missing for C as we do not have measurements of C for giants.
\label{fig:age_abund_relation_mstogiant}}
\end{figure*}

\begin{figure*}
\centering \includegraphics[width=0.99\textwidth]{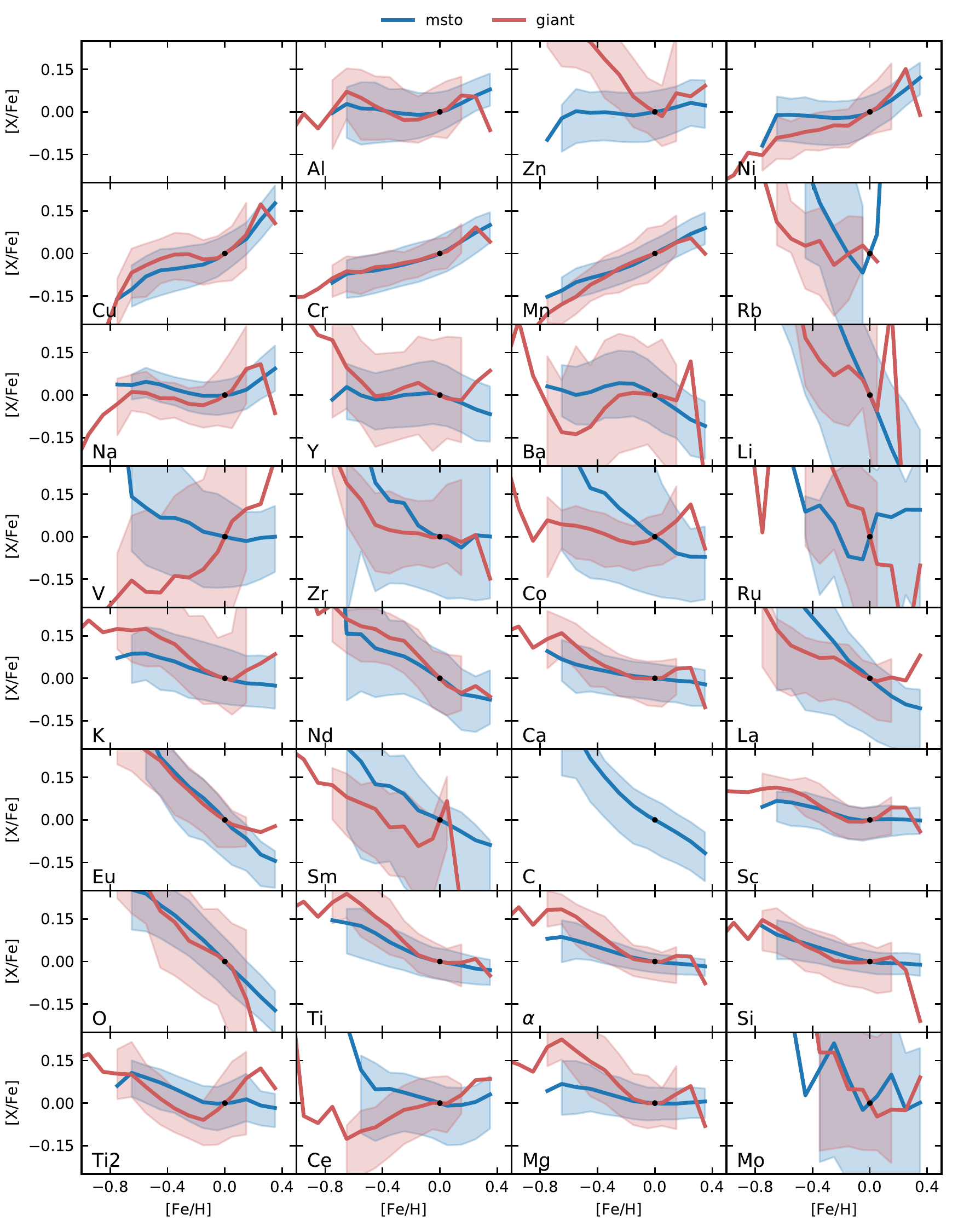}
\caption{Abundance of different elements as a function of [Fe/H] for MSTO and seismic giant stars with SNR$>40$. The solid lines shows the function $f_{\rm [Fe/H],X}$ as given by \autoref{equ:empirical_model}. The shaded region shows the 16 and 84 percentile spread around the relation (spread of [X/H]-$f_{\tau, {\rm X}}-f_{\odot, {\rm X}}$) for stars with SNR$>40$. The red profile is missing for C as we do not have measurements of C for giants.
\label{fig:feh_abund_relation_mstogiant_feh}}
\end{figure*}

\subsection{Comparison of abundance trends of MSTO stars with that of giants}
We have shown that the chemical abundances of an MSTO star can be predicted from its age and metallicity. In principle, the giants should also follow the same abundance-age-metallicity relations as the MSTO stars. However, in practise various factors can make the giants behave differently from that of MSTO stars, such as atomic diffusion \citep{2019A&A...627A.117L}, systematic surface gravity dependent effects in spectroscopic analyses, systematics in age estimates, abundance relations not being valid for all Galactocentric radii, $R$.

In \autoref{fig:age_abund_relation_mstogiant} we show $f_{\tau,{\rm X}}$ for the MSTO stars (blue curve) alongside that of giants (red curve) for different elements. The shaded region shows the 16 and 84 percentile spread around the relation (spread of [X/H]-$f_{\rm [Fe/H], X}-f_{\odot, {\rm X}}$) for stars with SNR$>40$. The trends of giants are in agreement with that of the MSTO stars; the blue and red lines track each other, and the blue and the red shaded regions show good overlap. Note, there are very few stars for which Rb, Li, Ru, C, and Mo can be measured in giants, hence we exclude them from further discussion in this section.  Among the elements that can be reliably measured in giants some elements show differences with respect to the MSTO stars. For elements having a non zero slope with age; such as Al, Zn, Y, Ba, Ca, Eu, Sc, O, Ti, Si, Ce, and Mg; at the old end, the giants seem to under predict the abundances if the slope is positive and over predict the abundances if the slope is negative. This could be due to larger uncertainties in age for giants, which means that an old age bin can have significant contamination from young stars.

In \autoref{fig:feh_abund_relation_mstogiant_feh} we show $f_{\rm [Fe/H], X}$ for MSTO stars (blue curve) alongside that of giants (red curve) for different elements. Generally, there is good agreement in trends between giants and MSTO stars, except for Zn ,V, Co, and Ce at the metal poor end. Zr, La, Ti, and Mg also show some differences at the metal poor end. This could be because most of these elements also have strong dependence on either SNR, \teff, or $\log g$.

\section{Discussions}

\subsection{Comparison with other studies}
\citet{2018ApJ...865...68B}, using solar twins (stars with \teff, $\log g$, and [Fe/H] similar to the Sun) showed that scatter around the abundance-age relation is small with values between 0.01 to 0.06 and the median being around 0.02 \citep[see also][]{2015A&A...579A..52N}. These studies represent one of the most precise abundance measurements (< 0.01 dex) due to the use of line-by-line differential analyses and the spectra having high resolution and wide wavelength coverage.  However, the sample size of the solar twin studies was small (79) and the data only explored stars with solar metallicity and had a limited span in age (0 to 8 Gyr) and Galactocentric radius $R$. Hence, it is not obvious that all Galactic stars should have a small scatter for a given age and metallicity. We explore abundances with significantly more stars that span a wide range in age, metallicity,  $R$ and $z$.  Our estimate of 0.033 for the abundance scatter of MSTO stars is in good agreement with the solar twin studies.

Analysis of abundance-age relations using spectra with high resolution and wide wavelength range was extended to a wider metallicity range by a number of studies \citep{2017MNRAS.465L.109F,2019A&A...624A..78D, 2020A&A...639A.127C}.  The results in general point to small abundance scatter for a given age and metallicity, although the scatter was not explicitly reported.  The focus of those studies was to identify elements that correlate with age and hence could be used as clocks to date the stars. The chemical abundances were assumed to be linear functions of age and metallicity, which is only valid for thin disc stars having low [$\alpha$/Fe].
\citet{2020A&A...639A.127C} found that for some elements, stars with Fe/H $> 0.1$ showed a slightly different slope for the age abundance relations as compared to stars with [Fe/H] $< 0.1$, with the element Y being a prime example.
We do not find any significant difference in the abundance-age relation for the metal rich and the metal poor stars; except for Ba, which showed a small difference. Our study has the advantage of using significantly more stars (50000 as compared to 560), however, our abundance precision is lower which can potentially erode small differences. It is not clear if the differences between our result for [Y/Fe] and that of \citet{2020A&A...639A.127C} is due to our lower abundance precision or if it is due to systematic differences between the spectroscopic analyses. In spite of the differences, their results are still consistent with the fact that abundances can be predicted to good accuracy from just age and metallicity, which is also one of our main conclusions.

\citet{2019ApJ...883..177N} performed a detailed analysis of the elemental abundance scatter for a large number of stars using the large scale spectroscopic survey APOGEE. They showed that the abundance scatter for a given age and metallicity is around 0.02 dex. Our abundance scatter is slightly higher (around 0.05 dex), but overall it is in good agreement with their results. We both arrive at the same conclusion that the abundance of most elements can be predicted from just age and metallicity.  We complement and extend the \citet{2019ApJ...883..177N} analysis in a number of ways. The abundance scatter in \citet{2019ApJ...883..177N} was computed for two different data sets, one having direct age estimates from asteroseismic information provided by Kepler and the other having age estimated purely from abundances using empirical relations. The data set with direct age estimates only had about 600 stars and was confined to the solar radius. The latter data set was larger (15000 stars) and explored stars well beyond the solar radius. Using giants with asteroseismic information from K2, we here extend the analysis of direct age measurements by having a larger number of stars (4000) and by extending beyond the solar radius. Additionally, we extend the \citet{2019ApJ...883..177N} analysis by also exploring $s$- and $r$-process elements that have different nucleosythetic origin. Finally, they restricted their analysis to stars with low [$\alpha$/Fe], which biases the sample to show a small scatter for the $\alpha$ elements, while we explore stars of all [$\alpha$/Fe].

\citet{2019ApJ...874..102W} showed that the median trends of [X/Mg] with [Mg/H] for two
populations, high [Mg/Fe] and low [Mg/Fe] are independent of Galactic location.
This can be interpreted as follows, [X/Fe] can be predicted from just [Mg/Fe] and [Fe/H]. This works because they analyzed elements that were produced by
SN Ia (EWD) and core-collapse SNe (EMS), and  [Mg/Fe] and [Fe/H] can
effectively account for these two processes. However, elements produced by $s$-process
cannot be predicted by just [Mg/Fe] and [Fe/H]. This is because, for age less than 10 Gyr,
the abundance of $s$-process elements have strong dependence on age, while
[Mg/Fe] and [Fe/H] have very little dependence on age (\autoref{fig:age_abund_relation_msto}).

\subsection{Physical interpretation of the abundance trends.}
\label{sec:physinterp}
Abundance trends with age and metallicity are different for different elements; some are flat, some have positive slopes, some have negative slopes, and some show non monotonic behavior(a rise accompanied by a fall or vice versa).  Trends for some of the elements can be explained by identifying the main nucleosynthetic process and the life time of the stars in which they are synthesized (production sites). Some of the main production sites of elements are, exploding massive star (EMS for example SNe II), exploding white dwarf (EWD for example SNe Ia) and AGB stars \citep{2020ApJ...900..179K}. EMS stars have the shortest life times and are the earliest to start the chemical enrichment of the Galaxy. EWD and AGB stars have long life times ranging from 0.5 to 20 Gyr and hence they enrich the ISM over longer time scales. The abundances of EMS elements are expected to increase with age, along with a sharp jump at around 10 Gyr due to time delay of about a Gyr between the onset of EWD and EMS \citep{2009MNRAS.396..203S}.  The abundances of EMS elements are expected to fall with increasing [Fe/H] as we measure abundances relative to Fe, about 50\% of which is produced by EWD. The abundances of EWD elements are expected to decrease mildly with age but increase with increasing [Fe/H]. This is because Fe is produced both in EMS and EWD. The abundances of AGB elements are expected to fall as age increases, as AGB stars have the longest life times and have a significant delay time to enrich the ISM. They are also expected to fall with increasing [Fe/H] as they are not produced along with Fe. Note, this is a very simplistic view of the chemical evolution in the Milky Way. In reality, nucleosynthetic yields of certain elements depend upon the metallicity and the progenitor mass of the exploding SNe.

Based on the simple model above, we now analyze the trends of different elements individually as shown in \autoref{fig:feh_abund_relation_msto}, \autoref{fig:age_abund_relation_msto}, and \autoref{fig:classification}). We focus mainly on the MSTO stars, which have larger sample sizes and higher abundance precision than the giants.  The elements Ni, Cr, Mn, and Cu have little variation with age but they increase strongly with [Fe/H]. The first three are iron peak elements.  Like Fe, they are produced both in EWD and EMS; an increase of [X/Fe] with [Fe/H] indicates a larger contribution from EWD compared to that for [Fe/H]. Cu is an odd Z element, a rise with [Fe/H] for [Fe/H] $> 0$, as seen here, is predicted by the Galactic chemical evolution model of \citet{2020ApJ...900..179K}. Y, Ba, and Zr are $s$-process elements synthesized in AGB stars, and as expected their abundances decrease with age and metallicity.  V and Co were also found to show a behaviour similar to $s$-process elements, but with much weaker dependence on age; however, they also show dependence with other stellar parameters like SNR, \teff\ and $\log g$.  Mg, Si, Sc, Ca, Ti, Eu, C, and O are all even-Z EMS elements and as expected their abundances increase with age and decrease with metallicity.

Some elements are known to be synthesized in EMS stars, however, their trends differ from that of typical EMS elements. For example, Al and Zn are EMS elements and their abundances increase with age as expected; however, unlike other EMS elements, their dependence on [Fe/H] is almost flat except for a small rise for [Fe/H] $> 0$. Na is an EMS element, but shows a non monotonic behavior with both age and [Fe/H]. Finally, K is also an EMS element and it decreases with [Fe/H], but it has a non monotonic behaviour with age.  The slightly peculiar behaviour of Na, Al, Cu, and K could be due to the fact that these are odd Z elements, and their yields are known to depend on the metallicity of their progenitors \citep{2020ApJ...900..179K}.

V, Zr, Co, Ce, Nd, and La show dependence on stellar parameters like SNR, \teff, and $\log g$, which is not yet understood. Hence, it is difficult to interpret their abundance trends with age and [Fe/H] in any meaningful way.  V and Co decrease with [Fe/H] in agreement with \citet{2020ApJ...900..179K}, but they also show a mild decrease with age, which may or not be expected.  Ce, Nd, and La behave like EMS elements (increase with age and decrease strongly with metallicity), but according to the nucleosynthetic periodic table of \citet{2020ApJ...900..179K}, they should behave like AGB elements--decrease with age and have a weak dependence on [Fe/H]. Ru, Rb, Sm, and Mo are not well measured by GALAH, hence we do not discuss their trends. Li is known to be sensitive to \teff\  and $\log g$, which makes it difficult to interpret its dependence on age and metallicity.

\subsection{Why are abundances a function of age and metallicity?}
Fundamentally, chemical enrichment at a given location in the Galaxy depends upon the infall of fresh gas (determines the number of H atoms), the current star formation rate (number density of EMS events that inject certain elements like O and Si) and the stellar number density (number density of EWD events that inject Fe). All of these processes have a  dependence on age and birth radius. This suggests that chemical abundances should also, on average, be a function of age and birth radius and should have little variation along the azimuth.  We know that the ISM currently has a negative radial metallicity gradient; as it is seen in Cepheid stars \citep{2013A&A...558A..31L}, which because of their youth have not scattered much from their place of birth.  Various other studies have also reported a radial metallicity gradient for stars close to the Galactic midplane \citep{2014AJ....147..116H,2020arXiv201102533S}.  Most chemical evolution models also predict metallicity to fall off with birth radius \citep{2009MNRAS.396..203S}.  It is likely that metallicity has been a monotonically decreasing function of radius for all ages.  If so then for a given age, the metallicity can be used to estimate the birth radius of a star \citep{2018MNRAS.481.1645M}.  Hence, the abundances of different elements should be a function of age and metallicity.

If elemental abundances are a function of age and birth radius, then the ISM should be azimuthally homogeneous. However, the extent to which the ISM is azimuthally homogeneous is not known. Stochastic chemical enrichment along with enhanced star formation along non-axisymmetric structures like spiral arms can generate abundance scatter at a given birth radius. The stochasticity can be understood as follows.  Stars are born in gas clumps inside giant molecular clouds ranging in mass from $10^4 {\rm M}_{\odot}$ to $10^6 {\rm M}_{\odot}$, which condense out of the ISM \citep{2019ARA&A..57..227K}.  As stars form and die they enrich the ISM around them.  Different parcels of the ISM can have distinct chemical abundance patterns depending upon the correlation length of the different production sites. If the correlation length is small a given gas parcel can have extra enrichment from a few production events, which will make its composition distinct from other parcels.  This makes the chemical enrichment stochastic. If enrichment from a single production event is large we expect large scatter in the chemical abundances of stars with a given age and [Fe/H]; and vice versa if the effect is small.  Below, we estimate the scatter for SNe driven chemical enrichment,
which is based on a calculation by \citet{2009nceg.book.....P}.  For a more general treatment of stochastic enrichment
see \citet{2018MNRAS.475.2236K}, where they estimate the scatter to be $\sim 0.1$ dex.

Ejecta from a single supernovae can extend upto 50 pc. Assuming a gas density of $0.041 {\rm M}_{\odot}{\rm pc}^{-3}$ from \citet{2015ApJ...814...13M} the ejecta can enrich about $2 \times 10^4 {\rm M}_{\odot}$ of the ISM. A core-collapse SNe ejects about 2 ${\rm M}_{\odot}$ of O and increases the mass fraction of O by $10^{-4}$. To bring the oxygen mass fraction of the ISM to that of the Sun (of $0.008$), 80 SNe events would be required. A Poissonian fluctuation of $\sqrt{80}$ leads to a scatter of 0.046 dex.

The scatter estimated above will be further reduced if additional mixing processes are at work, which are suggested by some theoretical studies.   Numerical simulations by \citet{2014Natur.513..523F} and \citet{2018MNRAS.481.5000A} suggest that gas clouds undergo homogenization due to turbulent mixing before they start forming stars, which lowers any existing scatter of abundances in the ISM.  \citet{1995A&A...294..432R} suggest that gas can be mixed on scales of i) $1<l/{\rm kpc}<10$ azimuthally by shear-induced turbulence from differential rotation, which can mix the gas on timescales of the order of $10^9$ yr, ii) $0.1<l/{\rm kpc}<1$ by cloud collision, expanding shells, and differential rotation on timescale of $10^8$ yr, and iii) $0.001<l/{\rm kpc}<0.1$ by turbulent diffusion on timescale of $2 \times 10^6$ yr.  We now compare this with the time scale for chemical enrichment.
Based on the local star formation rate, it would take 1.3 Gyr for 80 EMS events to occur in a volume of radius 50 pc.
Given the long time scale associated with stochastic enrichment, it is very likely that the mixing process will reduce the stochastic scatter of chemical abundances in the ISM.

\subsection{Implications for Galactic Archaeology}
The fact that the abundance of different elements have a small scatter when expressed in terms of age and [Fe/H] has significant implications for the field of Galactic archaeology.
It makes it easier to study certain aspects of Galactic evolution, like dynamical and chemical. However it makes it difficult to study certain other aspects, like  associating stars to their birth clusters. Below, we elaborate on implications for each of the above aspects in greater detail.

\subsubsection{Dynamical and chemical evolution}
Elemental abundances being a function of age and [Fe/H]
makes it possible to compute ages from abundances.
This is important because ages are hard to measure, except for a few specific type of stars, MSTO stars and giants with either asteroseismology or abundance measurements of C and N \citep{2016ApJ...823..114N,2016MNRAS.456.3655M, 2015MNRAS.453.1855M}.  We discuss the details of estimating ages from abundances in a companion paper (Hayden et al 2020). In short, using the empirical model from \autoref{sec:empirical}, given age, $\tau$, and metallicity, $F'$, we can predict the abundance, $X_{i,{\rm Model}}(\tau,F')$, of elements $i$ from $\tau$ and $F'$ with some intrinsic dispersion $\epsilon_{Xi}$.  Then using Bayes theorem the probability distribution of age with a prior term $p(\tau|F')$ can be written as \be
p(\tau|X_i,\sigma_{Xi},F,\sigma_F)= \int {\rm d} F' p(\tau|F')  \mathcal{N}\left(F|F',\sigma_{F}^2\right) \times \\
\prod_i \mathcal{N}\left(X_i|X_{i,{\rm Model}}(\tau,F'),\sigma_{Xi}^2+\epsilon_{Xi}^2\right).
\ee
where $\sigma_{Xi}$ and $\sigma_{F}$ are measurement uncertainties of $X_i$ and $F$ respectively, and $\mathcal N$ stands for a normal distribution conditioned on a given mean and variance.
If elemental abundances being a function of age and [Fe/H] is a consequence of
them being a function of age and birth radius $R_b$, we can use the abundances to also estimate the $R_b$. This is important because we do not know of
any other observable that can track the birth radius.

The ability to measure age and birth radius, $R_b$, is useful for Galactic archaeology in a number of ways. It is useful to understand the star formation history over the whole disc, $p(\tau,R_b)$. It is also useful to understand dynamical processes like radial migration and secular heating, which requires a multi-dimensional data set -- the distribution $p({\bf v},{\bf r}|\tau,R_b)$. For a system in equilibrium under a gravitational potential the six dimensional phase space  distribution function $f({\bf v},{\bf r})$ can be expressed in terms of the 3-dimensional action space distribution $f(J_{R},J_{\phi},J_z)$. Hence the effective dimensionality for exploring the distribution  $p({\bf v},{\bf r},\tau,R_b)$ is 5. To explore this 5 dimensional space comprehensively at least several million stars (50 stars in $10^5$ cells) scattered over a wide range of $R$ and $z$ will be required.  Upcoming large multi-object spectroscopic surveys like SDSS-V \citep{2019BAAS...51g.274K}, 4MOST \citep{2012SPIE.8446E..0TD}, WEAVE \citep{2012SPIE.8446E..0PD}, MOONS \citep{2014SPIE.9147E..0NC} are ideally suited to pursue such science.

If elemental abundances are a function age and metallicity (or $R_b$), then
the modelling of Galactic chemical evolution becomes simpler. This is because, one does not have to worry about the scatter of elemental abundances for stars having the same age and metallicity (or $R_b$).  Also, one does not have to worry about selection functions when comparing the observed  trends of abundances with age and metallicity with that of model predictions.
Observationally, fewer samples of stars are required to determine the abundance relations as a function of age and [Fe/H]. This means the relations can be determined using high precision studies of elemental abundances and ages. An example applying an abundance-age-metallicity relation to build a chemodynamical model with radial migration was presented in \citet{2020arXiv200503646S}. The model was successful in explaining the origin of the double sequence in the distribution of the stars in ([$\alpha$/Fe],[Fe/H]) plane and also how the distribution varies with different Galactic $(R,z)$ locations.

\subsubsection{Chemically tagging stars to their star forming aggregates}
An important use of abundances is to associate stars to their star forming aggregates, an idea known as chemical tagging \citep{2002ARA&A..40..487F}.
The most common star forming aggregates are giant molecular gas clouds and it is thought that they are chemically homogeneous but distinct from one another.
Gas clouds born at same time and Galactocentric radius, $R_b$, can have distinct abundances if the correlation lengths of elements ejected from different stars (such as SNe) are small. The abundance difference between stars will then tell us about the size of the gas clouds and the physics that determines the correlation length of different production sites \citep{2016AN....337..894B,2016ApJ...816...10T}.
Based on estimates of intrinsic abundance scatter presented here and in \citet{2019ApJ...883..177N}, an abundance precision of 0.03 or even less would be required, which makes it difficult to purse physics related to the mass distribution of clouds and the correlation length of the ejecta.
Another factor that makes chemical tagging difficult is that for the set of elements that we have analyzed, the abundances are strongly correlated with each other such that there are very few groups of elements that vary independent of each other. In future, it will be important to identify elements that track unique and new production sites.

\section{Summary \& Conclusions}
Using data from the GALAH+ survey, we have explored the age and metallicity dependence of the abundances of 27 elements for stars in the Galactic disc. We explored the relations with age measurements for about 50,000 MSTO stars  and 4,000 giants (with asteroseismic information from the NASA K2 mission), which is significantly more stars than in previous studies. To summarise, our main findings are:

\begin{itemize}
\item Abundances of most elements are, to a good approximation, a function of age and [Fe/H] with a mean intrinsic scatter of about 0.046 dex (over 23 elements) for giants and even less for MSTO stars (0.033 dex). The abundance relations are valid for both: MSTO stars and seismic giants, in the solar neighborhood and beyond the solar Galactocentric radius, for low and high [$\alpha$/Fe] stars, for young and old stars, and for metal rich and metal poor stars. The elements explored include those synthesized by different independent nucleosynthetic channels, like the $s-$process, the $r-$process and the $\alpha-$process.
\item Abundances vary smoothly with age and [Fe/H] for stars in the Galactic disc, with no special provision being required to accommodate the thick disc or high [$\alpha$/Fe] stars.
\item The dependence of abundance on age and [Fe/H] is largely additively separable. Chemical evolution models should help us to understand this.
\item Elements can be grouped based on the  direction of their abundance gradient in the (age,[Fe/H]) plane and different groups can be roughly associated with distinct nucleosynthetic production sites.
\item A handful of elements (Li, Co, Zr, La, V, and Sm for dwarfs and Ba, Y, K, and V for giants) cannot be predicted reliably from age and metallicity. They could be astrophysically interesting. However, they also have strong dependencies on other variables like \teff, $\log g$, and SNR and this needs to be investigated and understood in future.
\end{itemize}

The fact that the elemental abundances are a smooth function of age and metallicity and that they are additively separable in these variables is an interesting result. However, uncertainties in age and abundance could be partly responsible for this, suggesting further investigation with better precision on age and abundances.

Abundance trends for giants are similar to that of MSTO stars. Our giants explore stars beyond the solar radius, and hence these results suggest that the derived abundance relations are valid for most stars in the Galactic disc. We do see some weak trends for giants with stellar parameters like $R_g$, SNR, \teff, and $\log g$, which weakens the Galaxy-wide applicability of the relations. However, giants as compared to MSTO stars have larger uncertainty in age, [Fe/H], and abundances, which could be responsible for some of the trends with $R_g$, SNR, and \teff, and this warrants further investigation.

We discussed reasons that could lead to abundances being a function of age and [Fe/H]. The primary cause for this is the fact that the number density of events that enrich the ISM are on average a function of age and birth radius. This could lead to chemical abundances being a function of age and birth radius.  Assuming [Fe/H] tracks the birth radius for a given age,
it follows that abundances should be a function of age and birth radius.  We also explained the cause for the small intrinsic scatter in chemical abundances for a given age and [Fe/H] (or birth radius).  Using a simple stochastic enrichment scheme, we show that the intrinsic scatter in abundances for a given age and birth radius is expected to be small ($\sim 0.05$), and the scatter will be even smaller if there is additional mixing in the ISM.

The fact that the abundances of different elements are a function of age and [Fe/H] and that the intrinsic scatter about these relations is small has significant implications for the field of Galactic archaeology.   Firstly, it makes it possible to estimate the age and birth radius of stars using abundances.  With age and birth radius we can explore dynamical processes like scattering and radial migration, and also the star formation history in different radial zones.  Secondly, it simplifies the modelling of chemical evolution, which currently is plagued by various poorly understood processes.  However, the small intrinsic scatter in abundances make it difficult to associate stars to unique star forming clusters, because an abundance precision of less than 0.03 dex would be required.

MSTO stars, being intrinsically faint, are not easy to observe beyond the solar neighborhood. However, it is important to verify if the abundance relations for the MSTO stars are valid beyond the solar radius. This will require performing multi-object-spectroscopy on telescopes with large diameters.  Going forward, to better pin down the intrinsic scatter, the abundance precision of local MSTO samples should be improved and more stars will be required with spectra having better SNR, higher spectral resolution, and wider wavelength coverage. Also, improvements in spectroscopic analysis techniques would be required to reduce and understand the abundance systematics due to stellar parameters like \teff\ and $\log g$ and due to stellar types like dwarfs and giants.

\acknowledgments
SS is funded by a Senior Fellowship (University of Sydney), an ASTRO-3D Research Fellowship and JBH's Laureate Fellowship from the Australian Research Council (ARC).
JBH's research team is supported by an ARC Laureate Fellowship (FL140100278) and funds from ASTRO-3D. MJH is supported by an ASTRO-3D 4-year Research Fellowship.

The GALAH Survey is supported by the ARC Centre of Excellence for All Sky Astrophysics in 3 Dimensions (ASTRO 3D), through project CE170100013.
This work has made use of data acquired through the Australian Astronomical Observatory, under programs: GALAH, TESS-HERMES and K2-HERMES. We acknowledge the traditional owners of the land on which the AAT stands, the Gamilaraay people, and pay our respects to elders past and present.

This work has made use of data from the European Space Agency (ESA) mission
{\it Gaia} (\url{https://www.cosmos.esa.int/gaia}), processed by the {\it Gaia}
Data Processing and Analysis Consortium (DPAC,
\url{https://www.cosmos.esa.int/web/gaia/dpac/consortium}). Funding for the DPAC
has been provided by national institutions, in particular the institutions
participating in the {\it Gaia} Multilateral Agreement.

\bibliographystyle{yahapj}
\bibliography{}

\begin{thebibliography}{}
\providecommand\natexlab[1]{#1}
\providecommand\JournalTitle[1]{#1}

\bibitem[{{Andrews} {et~al.}(2017){Andrews}, {Weinberg}, {Sch{\"o}nrich}, \&
  {Johnson}}]{2017ApJ...835..224A}
{Andrews}, B.~H., {Weinberg}, D.~H., {Sch{\"o}nrich}, R., \& {Johnson}, J.~A.
  2017,
  \href{http://dx.doi.org/10.3847/1538-4357/835/2/224}{\JournalTitle{\apj},
  835, 224}

\bibitem[{{Armillotta} {et~al.}(2018){Armillotta}, {Krumholz}, \&
  {Fujimoto}}]{2018MNRAS.481.5000A}
{Armillotta}, L., {Krumholz}, M.~R., \& {Fujimoto}, Y. 2018,
  \href{http://dx.doi.org/10.1093/mnras/sty2625}{\JournalTitle{\mnras}, 481,
  5000}

\bibitem[{{Bedell} {et~al.}(2018){Bedell}, {Bean}, {Mel{\'e}ndez}, {Spina},
  {Ram{\'\i}rez}, {Asplund}, {Alves-Brito}, {dos Santos}, {Dreizler}, {Yong},
  {Monroe}, \& {Casagrande}}]{2018ApJ...865...68B}
{Bedell}, M., {Bean}, J.~L., {Mel{\'e}ndez}, J., {et~al.} 2018,
  \href{http://dx.doi.org/10.3847/1538-4357/aad908}{\JournalTitle{\apj}, 865,
  68}

\bibitem[{{Bland-Hawthorn} \& {Sharma}(2016)}]{2016AN....337..894B}
{Bland-Hawthorn}, J., \& {Sharma}, S. 2016,
  \href{http://dx.doi.org/10.1002/asna.201612393}{\JournalTitle{Astronomische
  Nachrichten}, 337, 894}

\bibitem[{{Buck}(2020)}]{2020MNRAS.491.5435B}
{Buck}, T. 2020,
  \href{http://dx.doi.org/10.1093/mnras/stz3289}{\JournalTitle{\mnras}, 491,
  5435}

\bibitem[{{Buder} {et~al.}(2020){Buder}, {Sharma}, {Kos}, {Amarsi},
  {Nordlander}, {Lind}, {Martell}, {Asplund}, {Bland-Hawthorn}, {Casey}, {De
  Silva}, {D'Orazi}, {Freeman}, {Hayden}, {Lewis}, {Lin}, {Schlesinger},
  {Simpson}, {Stello}, {Zucker}, {Zwitter}, {Beeson}, {Buck}, {Casagrande},
  {Clark}, {Cotar}, {Da Costa}, {de Grijs}, {Feuillet}, {Horner}, {Khanna},
  {Kafle}, {Liu}, {Montet}, {Nandakumar}, {Nataf}, {Ness}, {Spina}, {Traven},
  {Trepper-Garcia}, {Ting}, {Vogrincic}, {Wittenmyer}, {Zerjal}, \& {the GALAH
  collaboration}}]{2020arXiv201102505B}
{Buder}, S., {Sharma}, S., {Kos}, J., {et~al.} 2020, \JournalTitle{arXiv
  e-prints}, arXiv:2011.02505

\bibitem[{{Casali} {et~al.}(2020){Casali}, {Spina}, {Magrini}, {Karakas},
  {Kobayashi}, {Casey}, {Feltzing}, {Van der Swaelmen}, {Tsantaki},
  {Jofr{\'e}}, {Bragaglia}, {Feuillet}, {Bensby}, {Biazzo}, {Gonneau},
  {Tautvai{\v{s}}ien{\.{e}}}, {Baratella}, {Roccatagliata}, {Pancino}, {Sousa},
  {Adibekyan}, {Martell}, {Bayo}, {Jackson}, {Jeffries}, {Gilmore}, {Randich},
  {Alfaro}, {Koposov}, {Korn}, {Recio-Blanco}, {Smiljanic}, {Franciosini},
  {Hourihane}, {Monaco}, {Morbidelli}, {Sacco}, {Worley}, \&
  {Zaggia}}]{2020A&A...639A.127C}
{Casali}, G., {Spina}, L., {Magrini}, L., {et~al.} 2020,
  \href{http://dx.doi.org/10.1051/0004-6361/202038055}{\JournalTitle{\aap},
  639, A127}

\bibitem[{{Casey} {et~al.}(2019){Casey}, {Lattanzio}, {Aleti}, {Dowe},
  {Bland-Hawthorn}, {Buder}, {Lewis}, {Martell}, {Nordlander}, {Simpson},
  {Sharma}, \& {Zucker}}]{2019ApJ...887...73C}
{Casey}, A.~R., {Lattanzio}, J.~C., {Aleti}, A., {et~al.} 2019,
  \href{http://dx.doi.org/10.3847/1538-4357/ab4fea}{\JournalTitle{\apj}, 887,
  73}

\bibitem[{{Chiappini} {et~al.}(1997){Chiappini}, {Matteucci}, \&
  {Gratton}}]{1997ApJ...477..765C}
{Chiappini}, C., {Matteucci}, F., \& {Gratton}, R. 1997,
  \href{http://dx.doi.org/10.1086/303726}{\JournalTitle{\apj}, 477, 765}

\bibitem[{{Cirasuolo} {et~al.}(2014){Cirasuolo}, {Afonso}, {Carollo}, {Flores},
  {Maiolino}, {Oliva}, {Paltani}, {Vanzi}, {Evans}, {Abreu}, {Atkinson},
  {Babusiaux}, {Beard}, {Bauer}, {Bellazzini}, {Bender}, {Best}, {Bezawada},
  {Bonifacio}, {Bragaglia}, {Bryson}, {Busher}, {Cabral}, {Caputi}, {Centrone},
  {Chemla}, {Cimatti}, {Cioni}, {Clementini}, {Coelho}, {Crnojevic}, {Daddi},
  {Dunlop}, {Eales}, {Feltzing}, {Ferguson}, {Fisher}, {Fontana}, {Fynbo},
  {Garilli}, {Gilmore}, {Glauser}, {Guinouard}, {Hammer}, {Hastings}, {Hess},
  {Ivison}, {Jagourel}, {Jarvis}, {Kaper}, {Kauffman}, {Kitching}, {Lawrence},
  {Lee}, {Lemasle}, {Licausi}, {Lilly}, {Lorenzetti}, {Lunney}, {Maiolino},
  {Mannucci}, {McLure}, {Minniti}, {Montgomery}, {Muschielok}, {Nandra},
  {Navarro}, {Norberg}, {Oliver}, {Origlia}, {Padilla}, {Peacock}, {Pedichini},
  {Peng}, {Pentericci}, {Pragt}, {Puech}, {Randich}, {Rees}, {Renzini}, {Ryde},
  {Rodrigues}, {Roseboom}, {Royer}, {Saglia}, {Sanchez}, {Schiavon},
  {Schnetler}, {Sobral}, {Speziali}, {Sun}, {Stuik}, {Taylor}, {Taylor},
  {Todd}, {Tolstoy}, {Torres}, {Tosi}, {Vanzella}, {Venema}, {Vitali},
  {Wegner}, {Wells}, {Wild}, {Wright}, {Zamorani}, \&
  {Zoccali}}]{2014SPIE.9147E..0NC}
{Cirasuolo}, M., {Afonso}, J., {Carollo}, M., {et~al.} 2014,
  \href{http://dx.doi.org/10.1117/12.2056012}{in Society of Photo-Optical
  Instrumentation Engineers (SPIE) Conference Series, Vol. 9147, Ground-based
  and Airborne Instrumentation for Astronomy V, ed. S.~K. {Ramsay}, I.~S.
  {McLean}, \& H.~{Takami}}, 91470N

\bibitem[{{da Silva} {et~al.}(2012){da Silva}, {Porto de Mello}, {Milone}, {da
  Silva}, {Ribeiro}, \& {Rocha-Pinto}}]{2012A&A...542A..84D}
{da Silva}, R., {Porto de Mello}, G.~F., {Milone}, A.~C., {et~al.} 2012,
  \href{http://dx.doi.org/10.1051/0004-6361/201118751}{\JournalTitle{\aap},
  542, A84}

\bibitem[{{Dalton} {et~al.}(2012){Dalton}, {Trager}, {Abrams}, {Carter},
  {Bonifacio}, {Aguerri}, {MacIntosh}, {Evans}, {Lewis}, {Navarro}, {Agocs},
  {Dee}, {Rousset}, {Tosh}, {Middleton}, {Pragt}, {Terrett}, {Brock}, {Benn},
  {Verheijen}, {Cano Infantes}, {Bevil}, {Steele}, {Mottram}, {Bates},
  {Gribbin}, {Rey}, {Rodriguez}, {Delgado}, {Guinouard}, {Walton}, {Irwin},
  {Jagourel}, {Stuik}, {Gerlofsma}, {Roelfsma}, {Skillen}, {Ridings},
  {Balcells}, {Daban}, {Gouvret}, {Venema}, \& {Girard}}]{2012SPIE.8446E..0PD}
{Dalton}, G., {Trager}, S.~C., {Abrams}, D.~C., {et~al.} 2012,
  \href{http://dx.doi.org/10.1117/12.925950}{in Society of Photo-Optical
  Instrumentation Engineers (SPIE) Conference Series, Vol. 8446, Ground-based
  and Airborne Instrumentation for Astronomy IV, ed. I.~S. {McLean}, S.~K.
  {Ramsay}, \& H.~{Takami}}, 84460P

\bibitem[{{de Jong} {et~al.}(2012){de Jong}, {Bellido-Tirado}, {Chiappini},
  {Depagne}, {Haynes}, {Johl}, {Schnurr}, {Schwope}, {Walcher}, {Dionies},
  {Haynes}, {Kelz}, {Kitaura}, {Lamer}, {Minchev}, {M{\"u}ller}, {Nuza},
  {Olaya}, {Piffl}, {Popow}, {Steinmetz}, {Ural}, {Williams}, {Winkler},
  {Wisotzki}, {Ansorge}, {Banerji}, {Gonzalez Solares}, {Irwin}, {Kennicutt},
  {King}, {McMahon}, {Koposov}, {Parry}, {Sun}, {Walton}, {Finger}, {Iwert},
  {Krumpe}, {Lizon}, {Vincenzo}, {Amans}, {Bonifacio}, {Cohen}, {Francois},
  {Jagourel}, {Mignot}, {Royer}, {Sartoretti}, {Bender}, {Grupp}, {Hess},
  {Lang-Bardl}, {Muschielok}, {B{\"o}hringer}, {Boller}, {Bongiorno}, {Brusa},
  {Dwelly}, {Merloni}, {Nandra}, {Salvato}, {Pragt}, {Navarro}, {Gerlofsma},
  {Roelfsema}, {Dalton}, {Middleton}, {Tosh}, {Boeche}, {Caffau}, {Christlieb},
  {Grebel}, {Hansen}, {Koch}, {Ludwig}, {Quirrenbach}, {Sbordone}, {Seifert},
  {Thimm}, {Trifonov}, {Helmi}, {Trager}, {Feltzing}, {Korn}, \&
  {Boland}}]{2012SPIE.8446E..0TD}
{de Jong}, R.~S., {Bellido-Tirado}, O., {Chiappini}, C., {et~al.} 2012,
  \href{http://dx.doi.org/10.1117/12.926239}{in Society of Photo-Optical
  Instrumentation Engineers (SPIE) Conference Series, Vol. 8446, Ground-based
  and Airborne Instrumentation for Astronomy IV, ed. I.~S. {McLean}, S.~K.
  {Ramsay}, \& H.~{Takami}}, 84460T

\bibitem[{{Delgado Mena} {et~al.}(2019){Delgado Mena}, {Moya}, {Adibekyan},
  {Tsantaki}, {Gonz{\'a}lez Hern{\'a}ndez}, {Israelian}, {Davies}, {Chaplin},
  {Sousa}, {Ferreira}, \& {Santos}}]{2019A&A...624A..78D}
{Delgado Mena}, E., {Moya}, A., {Adibekyan}, V., {et~al.} 2019,
  \href{http://dx.doi.org/10.1051/0004-6361/201834783}{\JournalTitle{\aap},
  624, A78}

\bibitem[{{Dotter} {et~al.}(2017){Dotter}, {Conroy}, {Cargile}, \&
  {Asplund}}]{2017ApJ...840...99D}
{Dotter}, A., {Conroy}, C., {Cargile}, P., \& {Asplund}, M. 2017,
  \href{http://dx.doi.org/10.3847/1538-4357/aa6d10}{\JournalTitle{\apj}, 840,
  99}

\bibitem[{{Edvardsson} {et~al.}(1993){Edvardsson}, {Andersen}, {Gustafsson},
  {Lambert}, {Nissen}, \& {Tomkin}}]{1993A&A...275..101E}
{Edvardsson}, B., {Andersen}, J., {Gustafsson}, B., {et~al.} 1993,
  \JournalTitle{\aap}, 500, 391

\bibitem[{{Feltzing} {et~al.}(2017){Feltzing}, {Howes}, {McMillan}, \&
  {Stonkut{\.{e}}}}]{2017MNRAS.465L.109F}
{Feltzing}, S., {Howes}, L.~M., {McMillan}, P.~J., \& {Stonkut{\.{e}}}, E.
  2017, \href{http://dx.doi.org/10.1093/mnrasl/slw209}{\JournalTitle{\mnras},
  465, L109}

\bibitem[{{Feng} \& {Krumholz}(2014)}]{2014Natur.513..523F}
{Feng}, Y., \& {Krumholz}, M.~R. 2014,
  \href{http://dx.doi.org/10.1038/nature13662}{\JournalTitle{\nat}, 513, 523}

\bibitem[{{Freeman} \& {Bland-Hawthorn}(2002)}]{2002ARA&A..40..487F}
{Freeman}, K., \& {Bland-Hawthorn}, J. 2002,
  \href{http://dx.doi.org/10.1146/annurev.astro.40.060401.093840}{\JournalTitle{\araa},
  40, 487}

\bibitem[{{Gao} {et~al.}(2020){Gao}, {Lind}, {Amarsi}, {Buder},
  {Bland-Hawthorn}, {Campbell}, {Asplund}, {Casey}, {de Silva}, {Freeman},
  {Hayden}, {Lewis}, {Martell}, {Simpson}, {Sharma}, {Zucker}, {Zwitter},
  {Horner}, {Munari}, {Nordlander}, {Stello}, {Ting}, {Traven}, {Wittenmyer},
  \& {GALAH Collaboration}}]{2020MNRAS.497L..30G}
{Gao}, X., {Lind}, K., {Amarsi}, A.~M., {et~al.} 2020,
  \href{http://dx.doi.org/10.1093/mnrasl/slaa109}{\JournalTitle{\mnras}, 497,
  L30}

\bibitem[{{Hayden} {et~al.}(2014){Hayden}, {Holtzman}, {Bovy}, {Majewski},
  {Johnson}, {Allende Prieto}, {Beers}, {Cunha}, {Frinchaboy}, {Garc{\'\i}a
  P{\'e}rez}, {Girardi}, {Hearty}, {Lee}, {Nidever}, {Schiavon}, {Schlesinger},
  {Schneider}, {Schultheis}, {Shetrone}, {Smith}, {Zasowski}, {Bizyaev},
  {Feuillet}, {Hasselquist}, {Kinemuchi}, {Malanushenko}, {Malanushenko},
  {O'Connell}, {Pan}, \& {Stassun}}]{2014AJ....147..116H}
{Hayden}, M.~R., {Holtzman}, J.~A., {Bovy}, J., {et~al.} 2014,
  \href{http://dx.doi.org/10.1088/0004-6256/147/5/116}{\JournalTitle{\aj}, 147,
  116}

\bibitem[{{Kallinger} {et~al.}(2010){Kallinger}, {Mosser}, {Hekker}, {Huber},
  {Stello}, {Mathur}, {Basu}, {Bedding}, {Chaplin}, {De Ridder}, {Elsworth},
  {Frand sen}, {Garc{\'\i}a}, {Gruberbauer}, {Matthews}, {Borucki}, {Bruntt},
  {Christensen-Dalsgaard}, {Gilliland}, {Kjeldsen}, \&
  {Koch}}]{2010A&A...522A...1K}
{Kallinger}, T., {Mosser}, B., {Hekker}, S., {et~al.} 2010,
  \href{http://dx.doi.org/10.1051/0004-6361/201015263}{\JournalTitle{\aap},
  522, A1}

\bibitem[{{Kallinger} {et~al.}(2014){Kallinger}, {De Ridder}, {Hekker},
  {Mathur}, {Mosser}, {Gruberbauer}, {Garc{\'\i}a}, {Karoff}, \&
  {Ballot}}]{2014A&A...570A..41K}
{Kallinger}, T., {De Ridder}, J., {Hekker}, S., {et~al.} 2014,
  \href{http://dx.doi.org/10.1051/0004-6361/201424313}{\JournalTitle{\aap},
  570, A41}

\bibitem[{{Kobayashi} {et~al.}(2020){Kobayashi}, {Karakas}, \&
  {Lugaro}}]{2020ApJ...900..179K}
{Kobayashi}, C., {Karakas}, A.~I., \& {Lugaro}, M. 2020,
  \href{http://dx.doi.org/10.3847/1538-4357/abae65}{\JournalTitle{\apj}, 900,
  179}

\bibitem[{{Kollmeier} {et~al.}(2019){Kollmeier}, {Anderson}, {Blanc},
  {Blanton}, {Covey}, {Crane}, {Drory}, {Frinchaboy}, {Froning}, {Johnson},
  {Kneib}, {Kreckel}, {Merloni}, {Pellegrini}, {Pogge}, {Ramirez}, {Rix},
  {Sayres}, {S{\'a}nchez-Gallego}, {Shen}, {Tkachenko}, {Trump}, {Tuttle},
  {Weijmans}, {Zasowski}, {Barbuy}, {Beaton}, {Bergemann}, {Bochanski},
  {Brandt}, {Casey}, {Cherinka}, {Eracleous}, {Fan}, {Garc{\'\i}a}, {Green},
  {Hekker}, {Lane}, {Longa-Pe{\~n}a}, {Mathur}, {Meza}, {Minchev}, {Myers},
  {Nidever}, {Nitschelm}, {O'Connell}, {Price-Whelan}, {Raddick}, {Rossi},
  {Sankrit}, {Simon}, {Stutz}, {Ting}, {Trakhtenbrot}, {Weaver}, {Willmer}, \&
  {Weinberg}}]{2019BAAS...51g.274K}
{Kollmeier}, J., {Anderson}, S.~F., {Blanc}, G.~A., {et~al.} 2019, in Bulletin
  of the American Astronomical Society, Vol.~51, 274

\bibitem[{{Krumholz}(2014)}]{2014PhR...539...49K}
{Krumholz}, M.~R. 2014,
  \href{http://dx.doi.org/10.1016/j.physrep.2014.02.001}{\JournalTitle{\physrep},
  539, 49}

\bibitem[{{Krumholz} {et~al.}(2019){Krumholz}, {McKee}, \& {Bland
  -Hawthorn}}]{2019ARA&A..57..227K}
{Krumholz}, M.~R., {McKee}, C.~F., \& {Bland -Hawthorn}, J. 2019,
  \href{http://dx.doi.org/10.1146/annurev-astro-091918-104430}{\JournalTitle{\araa},
  57, 227}

\bibitem[{{Krumholz} \& {Ting}(2018)}]{2018MNRAS.475.2236K}
{Krumholz}, M.~R., \& {Ting}, Y.-S. 2018,
  \href{http://dx.doi.org/10.1093/mnras/stx3286}{\JournalTitle{\mnras}, 475,
  2236}

\bibitem[{{Kubryk} {et~al.}(2015){Kubryk}, {Prantzos}, \&
  {Athanassoula}}]{2015A&A...580A.126K}
{Kubryk}, M., {Prantzos}, N., \& {Athanassoula}, E. 2015,
  \href{http://dx.doi.org/10.1051/0004-6361/201424171}{\JournalTitle{\aap},
  580, A126}

\bibitem[{{Lemasle} {et~al.}(2013){Lemasle}, {Fran{\c{c}}ois}, {Genovali},
  {Kovtyukh}, {Bono}, {Inno}, {Laney}, {Kaper}, {Bergemann}, {Fabrizio},
  {Matsunaga}, {Pedicelli}, {Primas}, \& {Romaniello}}]{2013A&A...558A..31L}
{Lemasle}, B., {Fran{\c{c}}ois}, P., {Genovali}, K., {et~al.} 2013,
  \href{http://dx.doi.org/10.1051/0004-6361/201322115}{\JournalTitle{\aap},
  558, A31}

\bibitem[{{Lin} {et~al.}(2020){Lin}, {Asplund}, {Ting}, {Casagrand e}, {Buder},
  {Bland-Hawthorn}, {Casey}, {De Silva}, {D'Orazi}, {Freeman}, {Kos}, {Lind},
  {Martell}, {Sharma}, {Simpson}, {Zwitter}, {Zucker}, {Minchev},
  {{\v{C}}otar}, {Hayden}, {Horner}, {Lewis}, {Nordlander}, {Wyse}, \&
  {{\v{Z}}erjal}}]{2020MNRAS.491.2043L}
{Lin}, J., {Asplund}, M., {Ting}, Y.-S., {et~al.} 2020,
  \href{http://dx.doi.org/10.1093/mnras/stz3048}{\JournalTitle{\mnras}, 491,
  2043}

\bibitem[{{Liu} {et~al.}(2019){Liu}, {Asplund}, {Yong}, {Feltzing}, {Dotter},
  {Mel{\'e}ndez}, \& {Ram{\'\i}rez}}]{2019A&A...627A.117L}
{Liu}, F., {Asplund}, M., {Yong}, D., {et~al.} 2019,
  \href{http://dx.doi.org/10.1051/0004-6361/201935306}{\JournalTitle{\aap},
  627, A117}

\bibitem[{{Mackereth} {et~al.}(2018){Mackereth}, {Crain}, {Schiavon}, {Schaye},
  {Theuns}, \& {Schaller}}]{2018MNRAS.477.5072M}
{Mackereth}, J.~T., {Crain}, R.~A., {Schiavon}, R.~P., {et~al.} 2018,
  \href{http://dx.doi.org/10.1093/mnras/sty972}{\JournalTitle{\mnras}, 477,
  5072}

\bibitem[{{Marigo} {et~al.}(2017){Marigo}, {Girardi}, {Bressan}, {Rosenfield},
  {Aringer}, {Chen}, {Dussin}, {Nanni}, {Pastorelli}, {Rodrigues}, {Trabucchi},
  {Bladh}, {Dalcanton}, {Groenewegen}, {Montalb{\'a}n}, \&
  {Wood}}]{2017ApJ...835...77M}
{Marigo}, P., {Girardi}, L., {Bressan}, A., {et~al.} 2017,
  \href{http://dx.doi.org/10.3847/1538-4357/835/1/77}{\JournalTitle{\apj}, 835,
  77}

\bibitem[{{Martig} {et~al.}(2016){Martig}, {Fouesneau}, {Rix}, {Ness},
  {M{\'e}sz{\'a}ros}, {Garc{\'\i}a-Hern{\'a}ndez}, {Pinsonneault}, {Serenelli},
  {Silva Aguirre}, \& {Zamora}}]{2016MNRAS.456.3655M}
{Martig}, M., {Fouesneau}, M., {Rix}, H.-W., {et~al.} 2016,
  \href{http://dx.doi.org/10.1093/mnras/stv2830}{\JournalTitle{\mnras}, 456,
  3655}

\bibitem[{{Masseron} \& {Gilmore}(2015)}]{2015MNRAS.453.1855M}
{Masseron}, T., \& {Gilmore}, G. 2015,
  \href{http://dx.doi.org/10.1093/mnras/stv1731}{\JournalTitle{\mnras}, 453,
  1855}

\bibitem[{{McKee} {et~al.}(2015){McKee}, {Parravano}, \&
  {Hollenbach}}]{2015ApJ...814...13M}
{McKee}, C.~F., {Parravano}, A., \& {Hollenbach}, D.~J. 2015,
  \href{http://dx.doi.org/10.1088/0004-637X/814/1/13}{\JournalTitle{\apj}, 814,
  13}

\bibitem[{{Mel{\'e}ndez} {et~al.}(2009){Mel{\'e}ndez}, {Asplund}, {Gustafsson},
  \& {Yong}}]{2009ApJ...704L..66M}
{Mel{\'e}ndez}, J., {Asplund}, M., {Gustafsson}, B., \& {Yong}, D. 2009,
  \href{http://dx.doi.org/10.1088/0004-637X/704/1/L66}{\JournalTitle{\apjl},
  704, L66}

\bibitem[{{Minchev} {et~al.}(2013){Minchev}, {Chiappini}, \&
  {Martig}}]{2013A&A...558A...9M}
{Minchev}, I., {Chiappini}, C., \& {Martig}, M. 2013,
  \href{http://dx.doi.org/10.1051/0004-6361/201220189}{\JournalTitle{\aap},
  558, A9}

\bibitem[{{Minchev} {et~al.}(2018){Minchev}, {Anders}, {Recio-Blanco},
  {Chiappini}, {de Laverny}, {Queiroz}, {Steinmetz}, {Adibekyan}, {Carrillo},
  {Cescutti}, {Guiglion}, {Hayden}, {de Jong}, {Kordopatis}, {Majewski},
  {Martig}, \& {Santiago}}]{2018MNRAS.481.1645M}
{Minchev}, I., {Anders}, F., {Recio-Blanco}, A., {et~al.} 2018,
  \href{http://dx.doi.org/10.1093/mnras/sty2033}{\JournalTitle{\mnras}, 481,
  1645}

\bibitem[{{Ness} {et~al.}(2016){Ness}, {Hogg}, {Rix}, {Martig}, {Pinsonneault},
  \& {Ho}}]{2016ApJ...823..114N}
{Ness}, M., {Hogg}, D.~W., {Rix}, H.~W., {et~al.} 2016,
  \href{http://dx.doi.org/10.3847/0004-637X/823/2/114}{\JournalTitle{\apj},
  823, 114}

\bibitem[{{Ness} {et~al.}(2019){Ness}, {Johnston}, {Blancato}, {Rix}, {Beane},
  {Bird}, \& {Hawkins}}]{2019ApJ...883..177N}
{Ness}, M.~K., {Johnston}, K.~V., {Blancato}, K., {et~al.} 2019,
  \href{http://dx.doi.org/10.3847/1538-4357/ab3e3c}{\JournalTitle{\apj}, 883,
  177}

\bibitem[{{Nissen}(2015)}]{2015A&A...579A..52N}
{Nissen}, P.~E. 2015,
  \href{http://dx.doi.org/10.1051/0004-6361/201526269}{\JournalTitle{\aap},
  579, A52}

\bibitem[{{Nissen} {et~al.}(2017){Nissen}, {Silva Aguirre},
  {Christensen-Dalsgaard}, {Collet}, {Grundahl}, \&
  {Slumstrup}}]{2017A&A...608A.112N}
{Nissen}, P.~E., {Silva Aguirre}, V., {Christensen-Dalsgaard}, J., {et~al.}
  2017,
  \href{http://dx.doi.org/10.1051/0004-6361/201731845}{\JournalTitle{\aap},
  608, A112}

\bibitem[{{Pagel}(2009)}]{2009nceg.book.....P}
{Pagel}, B. E.~J. 2009, {Nucleosynthesis and Chemical Evolution of Galaxies}

\bibitem[{{Roy} \& {Kunth}(1995)}]{1995A&A...294..432R}
{Roy}, J.~R., \& {Kunth}, D. 1995, \JournalTitle{\aap}, 294, 432

\bibitem[{{Sch{\"o}nrich} \& {Binney}(2009)}]{2009MNRAS.396..203S}
{Sch{\"o}nrich}, R., \& {Binney}, J. 2009,
  \href{http://dx.doi.org/10.1111/j.1365-2966.2009.14750.x}{\JournalTitle{\mnras},
  396, 203}

\bibitem[{{Sharma} {et~al.}(2020){Sharma}, {Hayden}, \&
  {Bland-Hawthorn}}]{2020arXiv200503646S}
{Sharma}, S., {Hayden}, M.~R., \& {Bland-Hawthorn}, J. 2020,
  \JournalTitle{arXiv e-prints}, arXiv:2005.03646

\bibitem[{{Sharma} {et~al.}(2016){Sharma}, {Stello}, {Bland-Hawthorn}, {Huber},
  \& {Bedding}}]{2016ApJ...822...15S}
{Sharma}, S., {Stello}, D., {Bland-Hawthorn}, J., {Huber}, D., \& {Bedding},
  T.~R. 2016,
  \href{http://dx.doi.org/10.3847/0004-637X/822/1/15}{\JournalTitle{\apj}, 822,
  15}

\bibitem[{{Sharma} {et~al.}(2018){Sharma}, {Stello}, {Buder}, {Kos},
  {Bland-Hawthorn}, {Asplund}, {Duong}, {Lin}, {Lind}, {Ness}, {Huber},
  {Zwitter}, {Traven}, {Hon}, {Kafle}, {Khanna}, {Saddon}, {Anguiano}, {Casey},
  {Freeman}, {Martell}, {De Silva}, {Simpson}, {Wittenmyer}, \&
  {Zucker}}]{2018MNRAS.473.2004S}
{Sharma}, S., {Stello}, D., {Buder}, S., {et~al.} 2018,
  \href{http://dx.doi.org/10.1093/mnras/stx2582}{\JournalTitle{\mnras}, 473,
  2004}

\bibitem[{{Sharma} {et~al.}(2019){Sharma}, {Stello}, {Bland-Hawthorn},
  {Hayden}, {Zinn}, {Kallinger}, {Hon}, {Asplund}, {Buder}, {De Silva},
  {D'Orazi}, {Freeman}, {Kos}, {Lewis}, {Lin}, {Lind}, {Martell}, {Simpson},
  {Wittenmyer}, {Zucker}, {Zwitter}, {Bedding}, {Chen}, {Cotar}, {Esdaile},
  {Horner}, {Huber}, {Kafle}, {Khanna}, {Li}, {Ting}, {Nataf}, {Nordlander},
  {Saadon}, {Traven}, {Wright}, \& {Wyse}}]{2019MNRAS.490.5335S}
{Sharma}, S., {Stello}, D., {Bland-Hawthorn}, J., {et~al.} 2019,
  \href{http://dx.doi.org/10.1093/mnras/stz2861}{\JournalTitle{\mnras}, 490,
  5335}

\bibitem[{{Spina} {et~al.}(2016){Spina}, {Mel{\'e}ndez}, {Karakas},
  {Ram{\'\i}rez}, {Monroe}, {Asplund}, \& {Yong}}]{2016A&A...593A.125S}
{Spina}, L., {Mel{\'e}ndez}, J., {Karakas}, A.~I., {et~al.} 2016,
  \href{http://dx.doi.org/10.1051/0004-6361/201628557}{\JournalTitle{\aap},
  593, A125}

\bibitem[{{Spina} {et~al.}(2018){Spina}, {Mel{\'e}ndez}, {Karakas}, {dos
  Santos}, {Bedell}, {Asplund}, {Ram{\'\i}rez}, {Yong}, {Alves-Brito}, {Bean},
  \& {Dreizler}}]{2018MNRAS.474.2580S}
---. 2018,
  \href{http://dx.doi.org/10.1093/mnras/stx2938}{\JournalTitle{\mnras}, 474,
  2580}

\bibitem[{{Spina} {et~al.}(2020{\natexlab{a}}){Spina}, {Nordlander}, {Casey},
  {Bedell}, {D'Orazi}, {Mel{\'e}ndez}, {Karakas}, {Desidera}, {Baratella},
  {Yana Galarza}, \& {Casali}}]{2020ApJ...895...52S}
{Spina}, L., {Nordlander}, T., {Casey}, A.~R., {et~al.} 2020{\natexlab{a}},
  \href{http://dx.doi.org/10.3847/1538-4357/ab8bd7}{\JournalTitle{\apj}, 895,
  52}

\bibitem[{{Spina} {et~al.}(2020{\natexlab{b}}){Spina}, {Ting}, {De Silva},
  {Frankel}, {Sharma}, {Cantat-Gaudin}, {Joyce}, {Stello}, {Karakas},
  {Asplund}, {Nordlander}, {Casagrande}, {D'Orazi}, {Casey}, {Cottrell},
  {Tepper-Garc{\'\i}a}, {Baratella}, {Kos}, {{\v{C}}otar}, {Bland-Hawthorn},
  {Buder}, {Freeman}, {Hayden}, {Lewis}, {Lin}, {Lind}, {Martell},
  {Schlesinger}, {Simpson}, {Zucker}, \& {Zwitter}}]{2020arXiv201102533S}
{Spina}, L., {Ting}, Y.-S., {De Silva}, G.~M., {et~al.} 2020{\natexlab{b}},
  \JournalTitle{arXiv e-prints}, arXiv:2011.02533

\bibitem[{{Stello} {et~al.}(2015){Stello}, {Huber}, {Sharma}, {Johnson},
  {Lund}, {Handberg}, {Buzasi}, {Silva Aguirre}, {Chaplin}, {Miglio},
  {Pinsonneault}, {Basu}, {Bedding}, {Bland-Hawthorn}, {Casagrande}, {Davies},
  {Elsworth}, {Garcia}, {Mathur}, {Di Mauro}, {Mosser}, {Schneider},
  {Serenelli}, \& {Valentini}}]{2015ApJ...809L...3S}
{Stello}, D., {Huber}, D., {Sharma}, S., {et~al.} 2015,
  \href{http://dx.doi.org/10.1088/2041-8205/809/1/L3}{\JournalTitle{\apjl},
  809, L3}

\bibitem[{{Stello} {et~al.}(2017){Stello}, {Zinn}, {Elsworth}, {Garcia},
  {Kallinger}, {Mathur}, {Mosser}, {Sharma}, {Chaplin}, {Davies}, {Huber},
  {Jones}, {Miglio}, \& {Silva Aguirre}}]{2017ApJ...835...83S}
{Stello}, D., {Zinn}, J., {Elsworth}, Y., {et~al.} 2017,
  \href{http://dx.doi.org/10.3847/1538-4357/835/1/83}{\JournalTitle{\apj}, 835,
  83}

\bibitem[{{Ting} {et~al.}(2015){Ting}, {Conroy}, \&
  {Goodman}}]{2015ApJ...807..104T}
{Ting}, Y.-S., {Conroy}, C., \& {Goodman}, A. 2015,
  \href{http://dx.doi.org/10.1088/0004-637X/807/1/104}{\JournalTitle{\apj},
  807, 104}

\bibitem[{{Ting} {et~al.}(2016){Ting}, {Conroy}, \&
  {Rix}}]{2016ApJ...816...10T}
{Ting}, Y.-S., {Conroy}, C., \& {Rix}, H.-W. 2016,
  \href{http://dx.doi.org/10.3847/0004-637X/816/1/10}{\JournalTitle{\apj}, 816,
  10}

\bibitem[{{Weinberg} {et~al.}(2019){Weinberg}, {Holtzman}, {Hasselquist},
  {Bird}, {Johnson}, {Shetrone}, {Sobeck}, {Allende Prieto}, {Bizyaev},
  {Carrera}, {Cohen}, {Cunha}, {Ebelke}, {Fernandez-Trincado},
  {Garc{\'\i}a-Hern{\'a}ndez}, {Hayes}, {J{\"o}nsson}, {Lane}, {Majewski},
  {Malanushenko}, {M{\'e}sz{\'a}ros}, {Nidever}, {Nitschelm}, {Pan}, {Rix},
  {Rybizki}, {Schiavon}, {Schneider}, {Wilson}, \&
  {Zamora}}]{2019ApJ...874..102W}
{Weinberg}, D.~H., {Holtzman}, J.~A., {Hasselquist}, S., {et~al.} 2019,
  \href{http://dx.doi.org/10.3847/1538-4357/ab07c7}{\JournalTitle{\apj}, 874,
  102}

\bibitem[{{Yana Galarza} {et~al.}(2019){Yana Galarza}, {Mel{\'e}ndez},
  {Lorenzo-Oliveira}, {Valio}, {Reggiani}, {Carlos}, {Ponte}, {Spina},
  {Haywood}, \& {Gandolfi}}]{2019MNRAS.490L..86Y}
{Yana Galarza}, J., {Mel{\'e}ndez}, J., {Lorenzo-Oliveira}, D., {et~al.} 2019,
  \href{http://dx.doi.org/10.1093/mnrasl/slz153}{\JournalTitle{\mnras}, 490,
  L86}

\end{thebibliography}
\end{document}